\documentclass[12pt]{article} 

\usepackage{latexsym} 
\usepackage{amssymb}  
\usepackage{epsfig}       

\newcommand{\ep}{\varepsilon}

\newcommand{\<}{\langle} 
\renewcommand{\>}{\rangle} 

\newcommand{\beq}{\begin{equation}}
\newcommand{\eeq}{\end{equation}}
\newcommand{\ba}{\begin{array}}
\newcommand{\bea}{\begin{eqnarray}}
\newcommand{\ea}{\end{array}}
\newcommand{\eea}{\end{eqnarray}}

\renewcommand{\slash}{\!\!\!\!/\,}

\newcommand\comment[1]{ \hbox{[{\it Comment suppressed here.}\/]} }
\newcommand\hide[1]{}

\newcommand{\tr}{\hbox{tr}}
\newcommand{\Tr}{\hbox{Tr}}

\renewcommand{\Re}{ {\rm Re}\, }
\renewcommand{\Im}{ {\rm Im}\, }

\newcommand{\bx}{{\vec x}}
\newcommand{\by}{{\vec y}}

\newcommand{\bp}{{\vec p}}
\newcommand{\bq}{{\vec q}}

\newcommand{\bv}{{\vec v}}

\newcommand{\bA}{{\bf A}}
\newcommand{\bC}{{\bf C}}

\newcommand{\skipover}[1]{}

\newcommand{\C}{{\cal C}}

\makeatletter 

\def\appendix{\par                              
    \setcounter{section}{0}                     
    \setcounter{subsection}{0}
    \renewcommand{\theequation}{\Alph{section}.\arabic{equation}}
    \renewcommand{\thesection}{Appendix \Alph{section}}
}

\def\applabel#1{\@bsphack
  \protected@write\@auxout{}%
         {\string\newlabel{#1}{{\Alph{section}}{\thepage}}}%
  \@esphack}

\def\section{
\setcounter{equation}{0}        
\@startsection {section}{1}{\z@}{-3.5ex plus -1ex minus 
 -.2ex}{2.3ex plus .2ex}{\large\bf}}
\renewcommand{\theequation}{\arabic{section}.\arabic{equation}}

\def\subsection{\@startsection{subsection}{2}{\z@}{-3.25ex plus -1ex minus 
 -.2ex}{1.5ex plus .2ex}{\normalsize\bf}}

\def\subsubsection{\@startsection{subsubsection}{3}{\z@}{-3.25ex plus
 -1ex minus -.2ex}{1.5ex plus .2ex}{\normalsize}}

\makeatother   

\newsavebox{\eqlabel}

\makeatletter  
\newlength{\numblen}
\newsavebox{\eqnumb}
\def\@eqnnum{\savebox{\eqnumb}{\rm (\theequation)}%
\settowidth{\numblen}{\usebox{\eqnumb}}%
\makebox[\numblen][l]{\usebox{\eqnumb}~~~\usebox{\eqlabel}}}
\makeatother   

\newenvironment{equationwithlabel}[1]{ %
  \savebox{\eqlabel}{#1}
  \begin{equation}\label{#1} }{\end{equation}} 
\newcommand{\beql}[1]{\begin{equationwithlabel}{#1}}
\newcommand{\eeql}{\end{equationwithlabel}}

\begin{document}

\title{\bf Thermalization of fermionic\\
quantum fields\\[2.ex]}

\author{
J\"urgen Berges\thanks{email: j.berges@thphys.uni-heidelberg.de} 
$\,^a$,\addtocounter{footnote}{1}
Szabolcs Bors\'anyi\thanks{email: mazsx@cleopatra.elte.hu} $\,^{a,b}$ 
and
Julien Serreau\thanks{email: serreau@thphys.uni-heidelberg.de} $\,^a$
\\
[2.ex]
\normalsize{$^a$ Universit\"at Heidelberg, Institut f\"ur 
Theoretische Physik}\\
\normalsize{Philosophenweg 16, 69120 Heidelberg, Germany}\\
[1.ex]
\normalsize{$^b$ E{\"o}tv{\"o}s University, Department of Atomic Physics}\\
\normalsize{H-1117, Budapest, Hungary}\\
}

\newcommand{\preprintno}{
\normalsize HD-THEP-02-47
}
\date{
\preprintno}

\begin{titlepage}
\maketitle
\def\thepage{}          

\vspace*{-0.5cm}
\begin{abstract}
We solve the nonequilibrium dynamics of a $3+1$ dimensional theory 
with Dirac fermions coupled to scalars via a chirally invariant Yukawa 
interaction. The results are obtained from a systematic coupling expansion 
of the 2PI effective action to lowest non-trivial order, which includes 
scattering as well as memory and off-shell effects. The dynamics 
is solved numerically without further approximation, for different 
far-from-equilibrium initial conditions. The late-time behavior is 
demonstrated to be insensitive to the details of the initial conditions
and to be uniquely determined by the initial energy density.
Moreover, we show that at late time the system is very well
characterized by a thermal ensemble. In particular, we are able to 
observe the emergence of Fermi--Dirac and Bose--Einstein distributions  
from the nonequilibrium dynamics.
\end{abstract}

\end{titlepage}

\renewcommand{\thepage}{\arabic{page}}


\section{Introduction}
\label{soverview}

The abundance of experimental data on matter in extreme conditions from
relativistic heavy-ion collision experiments, as well as applications in
astrophysics and cosmology has lead to a strong increase of interest in the
dynamics of quantum fields out of equilibrium. Experimental indications 
of thermalization in collision experiments and the justification of current
predictions based on equilibrium thermodynamics, local equilibrium or
\mbox{(non--)linear} response pose major open questions for our theoretical
understanding~\cite{Braun-Munzinger:2001mh,Serreau:2002yr}. 
One way to resolve these questions is to try to understand quantitatively 
the far-from-equilibrium dynamics of quantum fields, without relying 
on the assumption of small departures from equilibrium, or on a possible 
separation of scales that forms the basis of effective kinetic 
descriptions \cite{Blaizot:2001nr}. In contrast to 
close-to-equilibrium approaches, the quantum-statistical fluctuations 
of the fields are not assumed to be described by a thermal ensemble. 
Moreover, one goes beyond the range of applicability of the usual gradient 
expansion and dilute-gas approximation.

In contrast to thermal equilibrium, which keeps no
information about the past, nonequilibrium dynamics poses 
an initial value problem: time-translation invariance is explicitly 
broken by the presence of the initial time, where the system has 
been prepared. The question of 
thermalization investigates how the system effectively looses the 
dependence on the details of the initial condition, and becomes 
approximately time-translation invariant at late times.
According to basic principles of equilibrium thermodynamics,
the thermal solution is universal in the sense that it is independent of the
details of the initial condition and is uniquely determined by the values of
the (conserved) energy density and of possible conserved
charges.\footnote{Here we consider closed systems without coupling to a heat
bath or external fields, which could provide sources or sinks of energy.} 
There are two distinct classes of universal behavior, corresponding to 
Bose-Einstein and Fermi-Dirac statistics respectively.

The description of the effective loss of initial conditions and subsequent
approach to thermal equilibrium in quantum field theory requires calculations
beyond so--called ``Gaussian'' (leading-order large--$N$, Hartree or mean-field
type) approximations \cite{gaussian1,gaussian2a,gaussian2b}. Similar to the
free-field theory limit, these approximations typically exhibit an infinite
number of additional conserved quantities which are not present in the
underlying interacting theory \cite{Cooper:1996ii,Aarts:2001wi}.  These
spurious constants of motion constrain the time evolution and lead to a
non-universal late-time behavior \cite{Berges:2001fi}. It has
recently been demonstrated in the context of scalar field theories, that the
approach to quantum thermal equilibrium can be described by going beyond these
approximations \cite{Berges:2000ur,Berges:2001fi}. In particular, this has been
achieved by using a systematic coupling expansion
\cite{Berges:2000ur} or a $1/N$ expansion to next-to-leading-order
\cite{Berges:2001fi,Aarts:2002dj} of the two-particle irreducible generating
functional for Green's functions, the so-called 2PI effective action
\cite{Baym,Cornwall:1974vz,Calzetta:1988cq,Ivanov:1998nv}.  In this
context, the corresponding equations of motion have been shown to
lead to a universal late time behavior, in the sense mentioned above, without
to have recourse to any kind of coarse-graining.

In this work, we study the far-from-equilibrium time evolution of relativistic
fermionic fields and their subsequent approach to thermal equilibrium. 
Nonequilibrium behavior of fermionic fields has been
previously addressed in several scenarios
\cite{gaussian2a,gaussian2b,Danielewicz:kk,Greene:2000ew,Joyce:2000ed,Lawrie:2000jg},
including the full dynamical problem with fermions coupled to inhomogeneous
{\em classical$\,$} bosonic fields \cite{Aarts:1998td}.  Here we go beyond
these approximations and compute the nonequilibrium evolution in a $3+1$
dimensional {\em quantum} field theory of Dirac fermions coupled to scalars in
a chirally invariant way.  As a consequence, we are able to study the approach
to quantum thermal equilibrium. For the considered
nonequilibrium initial conditions we find that the late-time behavior is
universal and characterized by Fermi--Dirac and Bose--Einstein distributions,
respectively.  The results are obtained from a systematic coupling expansion of
the 2PI effective action to lowest nontrivial (two-loop) order, 
which includes scattering
as well as memory and off-shell effects.  The nonequilibrium dynamics is 
solved numerically without further approximations. We emphasize that, given 
the limitations of a weak-coupling expansion, this is a first-principle 
calculation with no other input than the dynamics dictated by
the considered quantum field theory for given initial conditions.  

In Sect.~\ref{Sect2PIeffaction}, we review the 2PI effective action for 
fermions, which we use to derive exact time evolution equations for the 
spectral function and the statistical two-point function in 
Sect.~\ref{SectevolrhoF}. We discuss the Lorentz structure of our 
equations in Sect.~\ref{sec:lorentz} and exploit some symmetries in 
Sect.~\ref{sec:symmetries}. 
In Sect.~\ref{sec:model}, we specify to a chiral quark
model for which we solve the nonequilibrium dynamics. 
The initial conditions are discussed in Sect.~\ref{sec:initial}
and some details concerning the numerical implementation are
given in Sect.~\ref{sec:numerics}. The numerical results are 
presented and discussed in Sect.~\ref{Sectfarfromequilibrium} 
and Sect.~\ref{sec:statistics}. We attach two appendices discussing
in detail the quasiparticle picture we used to interprete some of
our results.

\section{2PI effective action for fermions}
\label{Sect2PIeffaction}

We consider first a purely fermionic quantum field theory with classical action
\beq
S = \int {\rm d}^4 x \Big(\bar{\psi}_i(x) [ i \partial\,\slash - m_f ] 
\psi_i(x) 
+ V(\bar{\psi},\psi) \Big) \, 
\label{fermact}
\eeq
for $i = 1,\ldots,N_f$ ``flavors'' of Dirac fermions $\psi_i$, a mass parameter
$m_f$ and an interaction term $V(\bar{\psi},\psi)$ to be specified below. Here
$\partial\,\slash \equiv \gamma^\mu \partial_\mu$, with Dirac matrices
$\gamma_\mu$ ($\mu =0,\ldots,3$). Summation over repeated indices and
contraction in Dirac space is implied.  All correlation functions of the
quantum theory can be obtained from the corresponding two-particle-irreducible
(2PI) effective action $\Gamma$. For the relevant case of a vanishing fermionic
``background'' field the 2PI effective action can be written as
\cite{Cornwall:1974vz}
\beq
\Gamma[D] = -i \Tr\ln D^{-1} 
          -i \Tr\, D_0^{-1} D
          + \Gamma_2[D] + {\rm const} \, .
\label{fermioneffact}
\eeq
The exact expression for the functional $\Gamma_2[D]$ contains all 2PI
diagrams with vertices described by $V(\bar{\psi},\psi)$ and propagator 
lines associated to the full connected two-point function
$D$. In coordinate space the trace $\Tr$ includes an integration over a  
closed time path $\C$ along the real axis \cite{Schwinger:1961qe},
as well as integration over spatial coordinates and summation over
flavor and Dirac indices. The free inverse propagator is given by
\beq
i D_{0,ij}^{-1} (x,y) 
= (i \partial\,\slash - m_f )\, \delta^4_{\cal C}(x-y)\, \delta_{ij} \, .
\eeq  
The equation of motion for $D$ in absence of external sources is 
obtained by extremizing the effective action \cite{Cornwall:1974vz}
\beq
\frac{\delta\Gamma[D]}{\delta D_{ij}(x,y)} = 0 \, .
\label{fermstat}
\eeq 
According to (\ref{fermioneffact}) one
can write (\ref{fermstat}) as an equation for the exact inverse propagator
\beq
D_{ij}^{-1}(x,y) = D_{0,ij}^{-1}(x,y) - \Sigma_{ij}(x,y;D) \, ,
\label{fermSD}
\eeq
with the proper self-energy 
\beq 
\Sigma_{ij}(x,y;D) \equiv -i \frac{\delta \Gamma_2[D]}{\delta D_{ji}(y,x)} \, .
\eeq  
Equation (\ref{fermSD}) can be rewritten in a form, which is suitable for 
initial value problems by convoluting it with $D$. One obtains the 
following time evolution equation for the propagator:
\beq
(i {\partial\,\slash}_{\!x} - m_f ) D_{ij}(x,y) 
- i \int_z \Sigma_{ik}(x,z;D) D_{kj}(z,y) =
i \delta_\C^4(x-y) \delta_{ij}  \,  ,
\label{evolutionC}
\eeq
where we employed the shorthand notation
$\int_z = \int_\C {\rm d}z^0 \int {\rm d}^3 z$.
Note that to keep the notation clear, we omit Dirac indices and 
reserve the Latin indices $i,j,k,\ldots$ to denote flavor.

\section{Exact evolution equations for the spectral 
and statistical components of the two-point function}
\label{SectevolrhoF}

To simplify physical interpretation we rewrite (\ref{evolutionC}) 
in terms of equivalent equations for the spectral function, which 
contains the information about the spectrum of the theory,
and the statistical two-point function. The latter will,
in particular, provide an effective description of occupation numbers.
For this we write for the time-ordered two-point function $D_{ij}(x,y)$
and proper self-energy $\Sigma_{ij}(x,y;D)$:\footnote{
If there is a local contribution to the proper self-energy,
we write 
\bea
\Sigma_{ij}(x,y;D) = - i\, 
\Sigma^{\rm (local)}(x;D)\,\delta_{\C}^4(x-y)\,\delta_{ij}
+ \Sigma_{ij}^{\rm (nonlocal)}(x,y;D) \, ,\nonumber
\eea
and the decomposition (\ref{sigbs}) is taken for 
$\Sigma^{\rm (nonlocal)}(x,y;D)$. In this case the local contribution
gives rise to an effective space-time dependent fermion mass term
$\sim m_f + \Sigma^{\rm (local)}(x;D)$.}  
\bea
D_{ij}(x,y)&=& \Theta_{\C}(x^0-y^0) \, D_{ij}^>(x,y) -
\Theta_{\C}(y^0-x^0) \, D_{ij}^<(x,y) \, ,
\label{Dbs}\\[0.2cm]
\Sigma_{ij}(x,y;D)&=&\Theta_{\C}(x^0-y^0) \, \Sigma_{ij}^>(x,y) -
\Theta_{\C}(y^0-x^0) \, \Sigma_{ij}^<(x,y) \label{sigbs} \, .
\eea
Note that for convenience we omit the explicit $D$--dependence in the
notation for $\Sigma_{ij}^{>,<}$. 
Inserting the above decompositions into the evolution
equation (\ref{evolutionC}), we
obtain evolution equations for the functions $D_{ij}^>(x,y)$ and
$D_{ij}^<(x,y)$ as well as the identity
\beq
\gamma^0 \left(D_{ij}^>(x,y) + D_{ij}^<(x,y) \right)|_{x^0=y^0} 
= \delta(\bx-\by) \delta_{ij} \, ,
\label{anticomrel}
\eeq
which corresponds to the anticommutation relation for fermionic field
operators. For later use we also note the hermiticity property 
\beq
\left(D_{ji}^>(y,x)\right)^\dagger
=\gamma^0 D_{ij}^>(x,y) \gamma^0 \, , 
\label{hermiticityD}
\eeq
and equivalently for $D_{ij}^<(x,y)$.
Note that here the hermitean conjugation denotes complex conjugation and 
taking the transpose in Dirac space only.

Similar to the discussion for scalar fields in
Refs.~\cite{Berges:2001fi,Aarts:2001qa} we introduce the spectral function
$\rho_{ij}(x,y)$ and the statistical propagator $F_{ij}(x,y)$ defined
as\footnote{Equivalently, one can decompose
$$
D_{ij}(x,y) = F_{ij}(x,y) - \frac{i}{2}\, \rho_{ij}(x,y) 
\left[\Theta_\C(x^0-y^0) - \Theta_\C(y^0-x^0) \right] \, . 
$$
}
\bea
\rho_{ij}(x,y) &=&i\, \Big(D_{ij}^>(x,y) + D_{ij}^<(x,y)\Big)\, , \\
F_{ij}(x,y) &=& \frac{1}{2}\, \Big(D_{ij}^>(x,y)-D_{ij}^<(x,y)\Big) \, . 
\eea
The corresponding components of the self-energy are given by\footnote{ Besides
the dynamical field degrees of freedom $D_{ij}$ we introduce quantities which
are functions of these fields. These functions are denoted by either boldface
or Greek letters.}
\bea
\bA_{ij}(x,y) &=&i\, \Big(\Sigma_{ij}^>(x,y) + \Sigma_{ij}^<(x,y)\Big)
\, , \\
\bC_{ij}(x,y) &=& \frac{1}{2}\, 
\Big(\Sigma_{ij}^>(x,y)-\Sigma_{ij}^<(x,y)\Big) \, . 
\eea
With (\ref{hermiticityD}) the two-point functions have the properties 
\bea
\left(\,\rho_{ji}(y,x)\,\right)^\dagger
&=& - \gamma^0 \rho_{ij}(x,y) \gamma^0 \, ,
\label{rhohermiticity}\\
\left(\,F_{ji}(y,x)\,\right)^\dagger
&=& \gamma^0 F_{ij}(x,y) \gamma^0 \, , \label{Fhermiticity} 
\eea
and equivalently for $\bA_{ij}(x,y)$ and $\bC_{ij}(x,y)$.

Using the above notations the evolution equation (\ref{evolutionC})
written for $\rho_{ij}(x,y)$ and $F_{ij}(x,y)$ are given by
\bea 
(i {\partial\,\slash}_{\!x} - m_f ) \rho_{ij} (x,y) &=& 
 \int_{y^0}^{x^0} {\rm d}z\,  \bA_{ik} (x,z)\rho_{kj} (z,y) \, , 
\label{rhoexact}\\
(i {\partial\,\slash}_{\!x} - m_f ) F_{ij}(x,y) &=& 
 \int_0^{x^0} {\rm d}z\, \bA_{ik}(x,z) F_{kj}(z,y)
\nonumber  \\ 
&-& \int_0^{y^0} {\rm d}z\, \bC_{ik}(x,z)\rho_{kj}(z,y) \, , 
\label{Fexact}
\eea
where we have taken the initial time to be zero, and
$\int_{y^0}^{x^0} {\rm d}z \equiv \int_{y^0}^{x^0} {\rm d}z^0 
\int {\rm d}^3 z$. For known self-energies the
equations (\ref{rhoexact}) and (\ref{Fexact}) are exact.  We note that the form
of their RHS is identical to the one for scalar 
fields \cite{Berges:2001fi,Aarts:2001qa}.
To solve the evolution equations one has to specify initial conditions for the
two-point functions, which is equivalent to specifying a Gaussian initial
density matrix\footnote{We emphasize that a Gaussian initial density matrix
only restricts the initial conditions or the ``experimental setup'' and
represents no approximation for the time evolution.  More general initial
conditions can be discussed using additional source terms in defining the
generating functional for Green's functions.}.
We note that the fermion anticommutation relation or (\ref{anticomrel})
uniquely specifies the initial condition for the spectral function:
\beq
\gamma^0 \rho_{ij}(x,y)|_{x^0=y^0} = i \delta(\bx-\by)\,\delta_{ij} \, .
\label{rhoinitial}
\eeq
Suitable nonequilibrium initial conditions for the statistical propagator
$F_{ij}(x,y)$ will be discussed below.

\section{Lorentz decomposition}
\label{sec:lorentz}

It is very useful to decompose the fields $\rho_{ij}(x,y)$ and $F_{ij}(x,y)$
into terms that have definite transformation properties under Lorentz
transformation. We will see below that, depending on the symmetry properties of
the initial state and interaction, a number of these terms remain identically
zero under the exact time evolution, which can dramatically simplify the
analysis. Using a standard basis and suppressing flavor indices we write
\beq
\rho = \rho_S + i \gamma_5 \rho_P + 
\gamma_\mu \rho_V^\mu + \gamma_\mu \gamma_5 \rho_A^\mu
+ \frac{1}{2} \sigma_{\mu\nu} \rho_T^{\mu\nu} \, ,
\label{rhodecomp}\\[0.1cm]
\eeq
where $\sigma_{\mu\nu} = \frac{i}{2}[\gamma_\mu,\gamma_\nu]$
and $\gamma_5=i\gamma^0\gamma^1\gamma^2\gamma^3$.
For given flavor indices the 16 (pseudo-)scalar, (pseudo-)vector and 
tensor components 
\bea
 \rho_S &=& \tilde{\tr}\, \rho \, , \nonumber\\
 \rho_P &=& - i\, \tilde{\tr}\, \gamma_5 \rho \, , \nonumber\\
 \rho_V^\mu &=& \tilde{\tr}\, \gamma^\mu \rho \, , \label{project}\\
 \rho_A^\mu &=& \tilde{\tr}\, \gamma_5 \gamma^\mu \rho \, , \nonumber\\
 \rho_T^{\mu\nu} &=& \tilde{\tr}\, \sigma^{\mu\nu} \rho  \, , \nonumber
\eea 
are complex two-point functions.  Here we have defined $\tilde{\tr} \equiv \frac{1}{4}
\tr$ where the trace acts in Dirac space. Equivalently, there are  16 complex
components for $F_{ij}$, $\bA_{ij}$ and $\bC_{ij}$ for given flavor indices $i,j$.
Using (\ref{rhohermiticity}) and (\ref{Fhermiticity}), one sees that they obey
\beq
\label{herm}
\rho^{(\Gamma)}_{ij}(x,y)= - \left(\rho^{(\Gamma)}_{ji}(y,x) \right)^*
\quad , \quad F^{(\Gamma)}_{ij}(x,y)= 
\left(F^{(\Gamma)}_{ji}(y,x)\right)^* \, ,
\eeq
where $\Gamma = \{S,P,V,A,T\}$. Inserting the above decomposition 
into the evolution equations (\ref{rhoexact}) and (\ref{Fexact})  
one obtains the respective equations for the various
components displayed in Eq.\ (\ref{project}). 
 
For a more detailed discussion, we first consider the LHS of the evolution
equations (\ref{rhoexact}) and (\ref{Fexact}). In fact, the approximation of a
vanishing RHS corresponds to the standard mean--field or Hartree--type
approaches frequently discussed in the literature \cite{gaussian1,gaussian2a}.
However, to discuss thermalization we have to go beyond such a ``Gaussian''
approximation: it is crucial to include direct scattering which is described
by the nonvanishing contributions from the RHS of the evolution equations.
Starting with the LHS of (\ref{rhoexact}) one finds, omitting flavor indices
(see also Ref.~\cite{Danielewicz:kk}):
\bea
 \tilde{\tr}\left[(i {\partial\,\slash} - m_f ) \rho \right] &=&
 \left(i \partial_\mu \rho_V^\mu\right) - m_f\, \rho_S \,\, ,
\label{rhoSL}\nonumber\\[0.3cm]
 -i\, \tilde{\tr}\left[\gamma_5 (i {\partial\,\slash} - m_f ) \rho \right] &=&
 - i \left(i \partial_\mu \rho_A^\mu\right) - m_f\, \rho_P \,\, ,
\label{rhoPL}\nonumber\\[0.3cm]
 \tilde{\tr}\left[\gamma^\mu (i {\partial\,\slash} - m_f ) \rho \right] &=&
 \left(i \partial^\mu \rho_S\right) 
 + i \left(i \partial_\nu \rho_T^{\nu\mu}\right) 
 - m_f\, \rho_V^\mu \,\, ,
\label{rhoVL}\\[0.1cm]
 \tilde{\tr}
\left[\gamma_5 \gamma^\mu (i {\partial\,\slash} - m_f ) \rho \right] &=&
 i \left(i \partial^\mu \rho_P\right) + \frac{1}{2}\, 
 \epsilon^{\mu\nu\gamma\delta} 
\left(i\partial_\nu \rho_{T,\gamma\delta}\right) 
 - m_f\, \rho_A^\mu \,\, ,
\label{rhoAL}\nonumber\\[0.1cm]
 \tilde{\tr}\left[\sigma^{\mu\nu} (i {\partial\,\slash} - m_f ) \rho \right] &=&
 -i \left(i\partial^\mu \rho_V^\nu - i \partial^\nu \rho_V^\mu\right)
 + \epsilon^{\mu\nu\gamma\delta} \left(i\partial_\gamma \rho_{A,\delta}\right)
 - m_f\, \rho_T^{\mu\nu} \, . \qquad
\label{rhoTL}\nonumber
\vspace*{0.1cm}
\eea
The corresponding expressions for the LHS of the evolution equation 
(\ref{Fexact}) for $F$ follow from
(\ref{rhoVL}) with the replacement $\rho \to F$. Considering now the various
component (\ref{project}) of the integrand on the RHS of Eq.\ (\ref{rhoexact}),
we find 
\bea
 \tilde{\tr}\left[\bA\, \rho \right] &=& 
 \bA_S\, \rho_S - \bA_P\, \rho_P + \bA_V^\mu\, \rho_{V,\mu}
 - \bA_A^\mu\, \rho_{A,\mu} \nonumber\\
 && + \frac{1}{2} \bA_T^{\mu\nu}\, \rho_{T,\mu\nu} 
 \,\, ,\quad 
\label{rhoSR}\\[0.1cm]
 -i\, \tilde{\tr}\left[\gamma_5 \bA\, \rho \right] &=& 
 \bA_S\, \rho_P + \bA_P\, \rho_S - i \bA_V^\mu\, \rho_{A,\mu}
 + i \bA_A^\mu\, \rho_{V,\mu} \nonumber\\
 && + \frac{1}{4}\, \epsilon^{\mu\nu\gamma\delta} \bA_{T,\mu\nu}\, 
 \rho_{T,\gamma\delta} \,\, ,
\label{rhoPR}\\[0.1cm]
 \tilde{\tr}\left[\gamma^\mu \bA\, \rho \right] &=& 
 \bA_S\, \rho_V^\mu + \bA_V^\mu\, \rho_S - i \bA_P\, \rho_A^\mu 
 + i \bA_A^\mu\, \rho_P + i \bA_{V,\nu}\, \rho_T^{\nu\mu}  \nonumber\\
 && + i \bA_T^{\mu\nu}\, \rho_{V,\nu}
 + \frac{1}{2} \epsilon^{\mu\nu\gamma\delta} \left( \bA_{A,\nu}\, 
 \rho_{T,\gamma\delta} + \bA_{T,\nu\gamma}\, 
 \rho_{A,\delta} \right) \, , \qquad
\label{rhoVR}\\[0.1cm]
 \tilde{\tr}\left[\gamma_5 \gamma^\mu \bA\, \rho \right] &=&
 \bA_S\, \rho_A^\mu + \bA_{A}^\mu\, \rho_S 
 - i \bA_P\, \rho_V^\mu + i \bA_V^\mu\, \rho_P
 + i \bA_{A,\nu}\, \rho_T^{\nu\mu}
  \nonumber\\
 && 
 + i \bA_T^{\mu\nu}\, \rho_{A,\nu} 
 + \frac{1}{2} \epsilon^{\mu\nu\gamma\delta} \left( \bA_{V,\nu}\, 
 \rho_{T,\gamma\delta} + \bA_{T,\nu\gamma}\, 
 \rho_{V,\delta}  \right) \, ,\quad
\label{rhoAR}\\[0.1cm]
 \tilde{\tr}\left[\sigma^{\mu\nu} \bA\, \rho \right] &=& 
 \bA_S\, \rho_T^{\mu\nu} + \bA_T^{\mu\nu}\, \rho_S 
 - \frac{1}{2}\,\epsilon^{\mu\nu\gamma\delta} 
 \left( \bA_P\, \rho_{T,\gamma\delta} + \bA_{T,\gamma\delta}\, 
\rho_P \right)
 \nonumber\\ 
&&
 - i \left(\bA_V^\mu\, \rho_V^\nu - \bA_V^\nu\, \rho_V^\mu \right) 
 + \epsilon^{\mu\nu\gamma\delta} \left( \bA_{V,\gamma}\, \rho_{A,\delta}
 - \bA_{A,\gamma}\, \rho_{V,\delta} \right)
 \nonumber\\[0.15cm] 
&&
 + i \left(\bA_A^\mu\, \rho_A^\nu - \bA_A^\nu\, \rho_A^\mu \right)
 + i \left(\bA_T^{\mu\gamma}\, {\rho_{T,\gamma}}^\nu 
 - \bA_T^{\nu\gamma}\, {\rho_{T,\gamma}}^\mu \right) .\qquad\,\,
\label{rhoTR}
\eea
With the above expressions one obtains the evolution equations for the various
Lorentz components in a straightforward way using (\ref{rhoexact}). We note
that the convolutions appearing on the RHS of the evolution equation
(\ref{Fexact}) for $F$ are of the same form than those computed above for
$\rho$. The respective RHS can be read off Eqs.\ (\ref{rhoSR})--(\ref{rhoTR}) 
by replacing $\rho \to F$ for the first term and $\bA \to \bC$ for the second 
term under the integrals of Eq.\ (\ref{Fexact}).  We have now all the relevant 
building blocks to discuss the most general case of nonequilibrium fermionic 
fields. However, this is often not necessary in practice due to the 
presence of symmetries, which require certain components to vanish 
identically.

\section{\label{sec:symmetries}Symmetries}

In the following, we will exploit symmetries of the action (\ref{fermact}) 
and of the initial conditions in order to simplify the fermionic 
evolution equations derived in the previous section. 

\vspace{.2cm} 
{\em Spatial translation invariance and isotropy:}
We will consider spatially homogeneous and isotropic initial conditions. In
this case it is convenient to work in Fourier space and we write
\beq
\rho(x,y) \equiv \rho(x^0,y^0;\bx-\by) = \int \frac{{\rm d}^3p}{(2\pi)^3}\,
e^{i \bp \cdot(\bx-\by)} \rho(x^0,y^0;\bp) \, ,
\eeq
and similarly for the other two-point functions. Moreover, isotropy implies 
a reduction of the number of independent two-point functions: e.g.~the vector
components of the spectral function can be written as 
\bea
 \rho_V^0(x^0,y^0;\bp) &=& \rho_V^0(x^0,y^0;p) \, , \nonumber \\
 \vec{\rho}_V(x^0,y^0;\bp) &=& \bv\, \rho_V(x^0,y^0;p) \, , \nonumber
\eea
where $p\equiv |\bp|$ and $\bv=\bp/p$. 

\vspace{.2cm}
{\em Parity:} The vector components $\rho_V^0(x^0,y^0;p)$ and $\rho_V(x^0,y^0;p)$ 
are unchanged under a parity transformation, whereas the corresponding axial-vector
components get a minus sign.  Therefore, parity together with rotational
invariance imply that
\beq
 \rho_A^0(x^0,y^0;p) = \rho_A(x^0,y^0;p) = 0 \, .
\eeq
The same is true for the axial-vector components of $F$, $\bA$ and $\bC$.
Parity also implies the pseudo-scalar components of the various two-point
functions to vanish.

\vspace{.2cm}
{\em $CP$--invariance:} For instance, under combined charge conjugation 
and parity transformation the vector component of $\rho$ transforms as
\bea
 \rho_V^0(x^0,y^0;p) &\longrightarrow& \rho_V^0(y^0,x^0;p) \, ,\nonumber\\
 \rho_V(x^0,y^0;p) &\longrightarrow& -\rho_V(y^0,x^0;p) \, ,\nonumber
\eea
and similarly for  $\bA_V^0$ and $\bA_V$.  The $F$--components transform as 
\bea
 F_V^0(x^0,y^0;p) &\longrightarrow& -F_V^0(y^0,x^0;p) \, , \nonumber\\
 F_V(x^0,y^0;p) &\longrightarrow& F_V(y^0,x^0;p) \, , \nonumber
\eea
and similarly for $\bC_V^0$ and $\bC_V$.  Combining this with the hermiticity
relations~(\ref{herm}), one obtains for these components that
\beq
\label{CP}
 \begin{array}{lll} 
  \Re \rho_V^0(x^0,y^0;p) &=& \Im \rho_V(x^0,y^0;p) = 0 \, , \\
  \Re F_V^0(x^0,y^0;p) &=& \Im F_V(x^0,y^0;p) = 0  \, ,\\
  \Re \bA_V^0(x^0,y^0;p) &=& \Im \bA_V(x^0,y^0;p) = 0  \, ,\\
  \Re \bC_V^0(x^0,y^0;p) &=& \Im \bC_V(x^0,y^0;p) = 0 \, ,
 \end{array}
\eeq
for all times $x^0$ and $y^0$ and all individual momentum modes.

\vspace{.2cm}
We stress that a nonequilibrium ensemble respecting a particular symmetry does
not imply that the individual ensemble members exhibit the same symmetry. For
instance, a spatially homogeneous ensemble can be build out of inhomogeneous
ensemble members and clearly includes the associated physics. The most
convenient choice of an ensemble is mainly dictated by the physical problem to
be investigated. Below we will study a ``chiral quark model'' with Dirac
fermions coupled to scalars in a chirally invariant way. Since in this paper we
will restrict the discussion of this model to the phase without spontaneous
breaking of chiral symmetry, it is useful to exploit this symmetry as well. 

\vspace{.2cm}
{\em Chiral symmetry:} 
The only components of the decomposition (\ref{rhodecomp}) allowed by
chiral symmetry are those which anticommute with $\gamma_5$. We therefore 
have
\beq
 \rho_S (x^0,y^0;\bp) = \rho_P (x^0,y^0;\bp) 
 = \rho_T^{\mu \nu} (x^0,y^0;\bp) = 0 \, ,
\eeq
and similarly for the corresponding components of $F$, $\bA$ and $\bC$. 
In particular, chiral symmetry forbids a mass term for fermions and we 
have $m_f\equiv0$.

\subsection{Equations of motion}

In conclusion, for the above symmetry properties we are left with only four
independent propagators: the two spectral functions $\rho_V^0$ and $\rho_V$ and
the two corresponding statistical functions $F_V^0$ and $F_V$. They are either
purely real or imaginary and have definite symmetry properties under the
exchange of their time arguments $x^0 \leftrightarrow y^0$. These properties as
well as the corresponding ones for the various components of the self-energy
are summarized below:
\begin{center}
\begin{tabular}{ll}
 $\rho_V^0$, $\bA_V^0$: & imaginary, symmetric; \\
 $\rho_V$, $\bA_V$: & real, antisymmetric;\\
 $F_V^0$, $\bC_V^0$: & imaginary, antisymmetric;\\
 $F_V$, $\bC_V$: & real, symmetric.
\end{tabular}
\end{center}
The exact evolution equations for the spectral functions 
read (cf.~Eq.~(\ref{rhoexact})):\footnote{We note that 
the following equations do not rely on the restrictions
(\ref{CP}) imposed by $CP$--invariance: they have the very same form 
for the case that all two-point functions are complex.}
\bea
\label{rhoV0eom}
\lefteqn{
i \frac{\partial}{\partial x^0}\, \rho_V^0(x^0,y^0;p)
= p\, {\rho}_V(x^0,y^0;p) } \\  
&+& \int_{y^0}^{x^0} {\rm d}z^0 \Big[
\bA_V^0(x^0,z^0;p)\, \rho_V^0(z^0,y^0;p) 
- {\bA}_V(x^0,z^0;p)\, {\rho}_V(z^0,y^0;p) \Big] \, ,
\nonumber
\\[0.2cm]
\label{rhoVeom}
\lefteqn{
i \frac{\partial}{\partial x^0}\, {\rho}_V(x^0,y^0;p)
= p\, \rho_V^0(x^0,y^0;p) } \\  
&+& \int_{y^0}^{x^0} {\rm d}z^0 \Big[
\bA_V^0(x^0,z^0;p)\, {\rho}_V(z^0,y^0;p) 
- {\bA}_V(x^0,z^0;p)\, \rho_V^0(z^0,y^0;p) \Big] \, .
\nonumber
\eea
Similarly, for the statistical two-point functions we obtain
(cf.~Eq.~(\ref{Fexact})):
\bea
\label{FV0eom}
\lefteqn{
i \frac{\partial}{\partial x^0}\, F_V^0(x^0,y^0;p)
= p\, {F}_V(x^0,y^0;p) } \\  
&+&  \int_0^{x^0} {\rm d}z^0 \Big[
\bA_V^0(x^0,z^0;p)\, F_V^0(z^0,y^0;p) 
- {\bA}_V(x^0,z^0;p)\, {F}_V(z^0,y^0;p) \Big] 
\nonumber\\
&-& \int_0^{y^0} {\rm d}z^0 \Big[
\bC_V^0(x^0,z^0;p)\, \rho_V^0(z^0,y^0;p) 
- {\bC}_V(x^0,z^0;p)\, {\rho}_V(z^0,y^0;p) \Big] \, ,
\nonumber\\
\nonumber\\[0.2cm]
\label{FVeom}
\lefteqn{
i \frac{\partial}{\partial x^0}\, {F}_V(x^0,y^0;p)
= p\, F_V^0(x^0,y^0;p) } \\  
&+& \int_0^{x^0} {\rm d}z^0 \Big[
\bA_V^0(x^0,z^0;p)\, {F}_V(z^0,y^0;p) 
- {\bA}_V(x^0,z^0;p)\, F_V^0(z^0,y^0;p) \Big] 
\nonumber \\
&-& \int_0^{y^0} {\rm d}z^0 \Big[
\bC_V^0(x^0,z^0;p)\, {\rho}_V(z^0,y^0;p) 
- {\bC}_V(x^0,z^0;p)\, \rho_V^0(z^0,y^0;p) \Big] \, .
\nonumber
\eea
The above equations are employed below to calculate the nonequilibrium
fermion dynamics in a chiral quark-meson model.  

\section{Chiral quark--meson model}
\label{sec:model}

As an application we consider a quantum field theory involving two fermion
flavors (``quarks'') coupled in a chirally invariant way to a scalar
$\sigma$--field and a triplet of pseudoscalar ``pions'' $\pi^a$
($a=1,2,3$). The classical action reads
\bea
S &=& \int {\rm d}^4 x \Big\{\bar{\psi} i \partial\,\slash \psi 
+\frac{1}{2}\left[\partial_\mu \sigma \partial^\mu \sigma
+ \partial_\mu \pi^a  \partial^\mu \pi^a \right] \nonumber\\
&& \qquad\quad\!\!\!  
+\, g \bar{\psi} \left[\sigma + i\gamma_5 \tau^a \pi^a \right] \psi
- V(\sigma^2 + \pi^2) \Big\} \, ,
\label{chiralfermact}
\eea
where $\pi^2\equiv \pi^a \pi^a$ and where $\tau^a$ denote the standard 
Pauli matrices. The above action is invariant under chiral 
$SU_L(2)\times SU_R(2)$ transformations.  
For a quartic scalar self-interaction 
$\sim \left(\sigma^2+ \pi^a\pi^a\right)^2$,
this model corresponds to the well known linear $\sigma$--model 
\cite{sigma}, which has been extensively studied in thermal equilibrium 
in the literature using various 
approximations \cite{sigma2}. For simplicity we consider here a 
purely quadratic 
scalar potential:  
\beq
 V  = \frac{1}{2} m^2_0 (\sigma^2 + \pi^2) \, ,
\label{potential}
\eeq
which is sufficient to study thermalization in this model.  We note that
this theory has the same universal properties than the corresponding linear
$\sigma$-model. Extending our study to take into account quartic
self-interaction is straightforward. It gives additional contributions to the
scalar self-energies, Eqs.~(\ref{selfscalarrho}) and (\ref{selfscalarF}) below.
These contributions can be found in Refs.~\cite{Berges:2001fi,Aarts:2002dj} and
we will point out the respective changes below.

\subsection{Equations of motion for the scalar field}

The 2PI effective action for nonequilibrium scalar fields has been 
extensively studied in the literature \cite{Calzetta:1988cq,Ivanov:1998nv,
Berges:2000ur,Berges:2001fi,Aarts:2002dj}. 
Here, we briefly recall the main features of the scalar sector and 
stress those aspects which are relevant for the present paper 
(for details see e.g.\ Refs.~\cite{Berges:2001fi,Aarts:2002dj}). 
For the model considered here, 
Eq.\ (\ref{chiralfermact}), the 2PI effective action is a functional 
of fermionic propagators as well as scalar propagators.\footnote{As 
emphasized in the previous section, we do not consider the possibility of a 
broken symmetry in this paper. Therefore we restrict the discussion 
to a vanishing scalar 
field expectation value.} The scalar fields form
an $O(4)$-vector $\phi_A(x) \equiv (\sigma(x), \vec{\pi}(x)\,)$ and
we denote the full scalar propagator by $\mathcal{G}_{AB}(x,y)$ 
with $A,B=0,\ldots,3$. 
The 2PI effective action (\ref{fermioneffact}), 
augmented by the scalar sector,
reads
\beq
 \Gamma[\mathcal{G},D] = \frac{i}{2} \Tr\ln \mathcal{G}^{-1} 
 + \frac{i}{2} \Tr \mathcal{G}_0^{-1} \mathcal{G}
 -i \Tr\ln D^{-1} -i \Tr\, D_0^{-1} D
 + \Gamma_2[\mathcal{G},D] + {\rm const} \, ,
\label{totaleffact}
\eeq
with the free scalar inverse propagator 
\beq
\label{classscalar}
 i \mathcal{G}_{0,AB}^{-1} (x,y) = -(\square_x + m^2_0 )\, 
 \delta^{4}_{\cal C}(x-y) \, \delta_{AB} \, .
\eeq
Similar to the fermionic case discussed above, the equation of motion 
for the scalar two-point function is
obtained by minimizing the 2PI effective action with respect to 
$G$, and the proper self-energy is given  
by~\cite{Cornwall:1974vz,Berges:2001fi,Aarts:2002dj}
\beq
\label{scalarself}
 \Sigma_{AB} (x,y) = 2i\,
 \frac{\delta\Gamma_2[\mathcal{G},D]}{\delta \mathcal{G}_{BA} (y,x)} \,.
\eeq
Under chiral transformations, the matrix $\mathcal{G}\rightarrow 
\mathcal{R} \, \mathcal{G} \,\mathcal{R}^\dagger$, where $\mathcal{R}$ is 
an $O(4)$ rotation. Without
loss of generality, because of chiral symmetry the effective action
$\Gamma[\mathcal{G},D]$ and its functional derivatives can be evaluated
for $\mathcal{G}$ taken to be the unit matrix in $O(4)$--space: 
\beq
\label{scalarsym}
 \mathcal{G}_{AB} (x,y)=G_\phi (x,y) \, \delta_{AB}\,.
\eeq 
The same holds for the corresponding self-energy (\ref{scalarself}). 
Similarly, in the fermionic sector, because of chiral symmetry the most
general fermion two-point function can be taken to be proportional to 
unity in flavor space: 
\beq
\label{fermionsym}
 D_{ij} (x,y)=D(x,y) \,\delta_{ij}\,.
\eeq 
Similar to the discussion in Sect.~\ref{SectevolrhoF}, 
the scalar spectral and statistical two-point functions are defined 
as~\cite{Berges:2001fi,Aarts:2001qa}
\beq
 G_\phi (x,y) = F_\phi (x,y) -\frac{i}{2} \rho_\phi (x,y)
 \left[\Theta_\C(x^0-y^0) - \Theta_\C(y^0-x^0) \right] \, ,
\eeq
and equivalently for the spectral and statistical self-energies
$\Sigma_\phi^{\rho}$ and $\Sigma_\phi^{F}$. These are all real functions and
$F$--like components are symmetric under the exchange of $x$ and $y$, whereas
the $\rho$--like components are antisymmetric.  The equal-time commutation
relation of two scalar field operators 
implies~\cite{Berges:2001fi,Aarts:2001qa}
\beq
\label{initialrhoscalar}
 \rho_\phi (x,y)|_{x^0=y^0} = 0, \;\;\;\;
 \partial_{x^0}\rho_\phi(x,y)|_{x^0=y^0} = \delta(\bx-\by)\, ,
\eeq 
which uniquely specifies the initial conditions for the spectral
function. Initial conditions for the statistical two-point
function will be discussed below.
Finally, the equations of motion for the scalar propagators 
read~\cite{Berges:2001fi,Aarts:2001qa}:
\bea 
\left[ \partial_{x^0}^2  + {\vec{p}\,}^2 + m^2_0 \right] 
\rho_\phi (x^0,y^0;\vec{p}) = 
  -\int_{y^0}^{x^0} {\rm d}z^0\,  
 \Sigma_\phi^{\rho} (x^0,z^0;\vec{p})\, \rho_\phi (z^0,y^0;\vec{p})\,. &&
 \nonumber\\
\label{rhoscalar}
\\[0.3cm]
 \left[ \partial_{x^0}^2 + {\vec{p}\,}^2 + m^2_0 \right] 
F_\phi(x^0,y^0;\vec{p}) = 
 - \int_0^{x^0} {\rm d}z^0\, \Sigma_\phi^{\rho}(x^0,z^0;\vec{p})\, 
 F_\phi(z^0,y^0;\vec{p})
\nonumber && \\ 
 + \int_0^{y^0} {\rm d}z^0\, \Sigma_\phi^{F}(x^0,z^0;\vec{p})\, 
 \rho_\phi(z^0,y^0;\vec{p}),&&\nonumber\\
\label{Fscalar}
\eea
where we have assumed spatially homogeneous initial conditions and Fourier
transformed with respect to spatial coordinates.
For known self-energies these are the exact equations for the 
theory described by the
classical action (\ref{chiralfermact}) together with (\ref{potential}). In the
presence of scalar self-interactions, the self-energy receives in
particular a local contribution 
$-i \Sigma_\phi^{\rm (local)}(x)\,\delta_\C^{4} (x-y)$, 
which amounts to a shift of the bare mass squared appearing in 
the above equations. 
In that case, the exact equations of motion have the 
same form as above, with the replacement 
$m_0^2 \rightarrow M^2(x) = m_0^2 + \Sigma_\phi^{\rm (local)} (x)$ 
\cite{Berges:2001fi,Aarts:2002dj}.

\subsection{2PI coupling expansion}

\begin{figure}[t]
\begin{center}
\epsfig{file=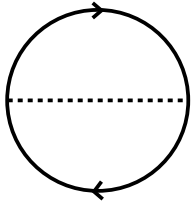,width=2.5cm}
\end{center}
\caption{
Two-loop contribution to $\Gamma_2[\mathcal{G},D]$.  The solid and dashed lines
represent the full fermion ($D$) and boson ($\mathcal{G}$)
propagators, respectively.}
\label{twoloopfig}
\end{figure}

We consider a systematic coupling expansion 
of the 2PI effective action (\ref{totaleffact}) 
to lowest non-trivial order, which includes 
scattering as well as memory and off-shell effects, such that
thermalization can be studied. The first non-trivial order in a coupling
expansion corresponds to the two-loop contribution depicted in 
Fig.~\ref{twoloopfig} (cf.~also the discussion in 
Ref.~\cite{Kainulainen:2002sw}).\footnote{We note that
this approximation can also be related to a nonperturbative
$1/N_f$ expansion at next-to-leading order.} 
Using (\ref{scalarsym}) and (\ref{fermionsym}), one
can express the two-loop contribution to $\Gamma_2$ directly in terms of
$G_\phi$ and~$D$. We obtain for the chirally symmetric theory:
\beq
 \Gamma_2^{\rm (2-loop)} [\mathcal{G},D]
 =-ig^2\frac{N_f N_s}{2}\int_\C {\rm d}^4x \,{\rm d}^4y \,
 \,\tr [D(x,y)\,D(y,x)]\,G_\phi(x,y) \, ,
\eeq
where $N_f=2$ is the number of fermion flavors and $N_s=4$ is the number of
scalar components. From there, it is straightforward to compute 
the spectral and statistical components of the two-loop 
self-energies\footnote{The relevant 
self-energies can be obtained with
$$
 2i\,\frac{\delta \Gamma_2}{\delta G_\phi} 
 = 2i\,\frac{\delta \Gamma_2}{\delta \mathcal{G}_{AB}}\,
 \frac{\delta \mathcal{G}_{AB}}{\delta G_\phi} 
 = \delta_{AB} \, \Sigma_{BA} = N_s \, \Sigma_\phi \, ,
$$
and similarly for the fermionic self energies $\Sigma_{ij}=\Sigma \,
\delta_{ij}$:
$$
 -i \frac{\delta \Gamma_2}{\delta D}= N_f \Sigma \, .
$$
}.  
We obtain for the scalar self-energies
\bea
\label{selfscalarrho}
 \Sigma_\phi^\rho (x^0,y^0;\bp) = -8 g^2 N_f\int \frac{{\rm d}^3 q}{(2\pi)^3}\,
 \rho_V^\mu(x^0,y^0;\bq)\,F_{V,\mu}(x^0,y^0;\bp-\bq) \, ,&&\\
\label{selfscalarF}
 \Sigma_\phi^F (x^0,y^0;\bp) = -4 g^2 N_f\int \frac{{\rm d}^3 q}{(2\pi)^3}\,
 \Big[F_V^\mu(x^0,y^0;\bq)\,F_{V,\mu}(x^0,y^0;\bp-\bq) &&\nonumber\\
 - \frac{1}{4} \rho_V^\mu(x^0,y^0;\bq)\,\rho_{V,\mu}(x^0,y^0;\bp-\bq) 
 \Big] \, ,&&
\eea
and for the fermion self-energies
\bea
\bA_V^\mu(x^0,y^0;\bp) = - g^2 N_s\int \frac{{\rm d}^3 q}{(2\pi)^3}\, \Big[
F_V^\mu(x^0,y^0;\bq) \,\rho_\phi(x^0,y^0;\bp-\bq) &&\nonumber\\
+ \rho_V^\mu(x^0,y^0;\bq)\, F_\phi(x^0,y^0;\bp-\bq)\Big] \, ,&&
\\[0.2cm] 
\bC_V^\mu(x^0,y^0;\bp) = - g^2 N_s\int \frac{{\rm d}^3q}{(2\pi)^3}\, \Big[
F_V^\mu(x^0,y^0;\bq) \,F_\phi(x^0,y^0;\bp-\bq) &&\nonumber\\
-\frac{1}{4} \rho_V^\mu(x^0,y^0;\bq)\, \rho_\phi(x^0,y^0;\bp-\bq)\Big]\, .&&
\label{selffermionF}
\eea
Finally, we note that for the case of a non-vanishing scalar
self-interaction, the only additional two-loop contribution gives 
rise to a local mass shift as described above. In particular, at this 
order there is no additional contribution to the direct scattering part,
i.e.\ to the RHS of Eqs.\ (\ref{rhoscalar}) and 
(\ref{Fscalar}) relevant for thermalization.

\section{Initial conditions}
\label{sec:initial}

The time evolution for the fermions is described by first-order 
\mbox{(integro-)}differential equations for $F$ and $\rho$: Eqs.\ 
(\ref{rhoV0eom})--(\ref{FVeom}). As pointed out above, the 
initial fermion spectral function is completely 
determined by the equal-time anticommutation relation of fermionic 
field operators (cf.~Eq.~(\ref{rhoinitial})). 
To uniquely specify the time evolution for $F$ we have to set the initial
conditions. The most general (Gaussian) initial conditions for $F$ respecting
spatial homogeneity, isotropy, parity, charge conjugation and chiral symmetry 
can be written as 
\bea
 F_V(t,t',p)|_{t=t'=0} &=& \frac{1}{2}- n_0^{f}(p) \, ,
\label{initialFV}\\
 F_V^0(t,t',p)|_{t=t'=0} &=& 0
\label{initialFV0} \, .
\eea
Here $n_0^{f}(p)$ denotes the initial particle number distribution, whose
values can range between 0 and 1 (the definition of effective particle number
distribution in terms of equal-time two-point functions is detailed in
\ref{sec:quasiparticle}). At late times, when thermal equilibrium 
is approached, this will lead to a canonical description with zero 
chemical potential.\footnote{In particular, for nonzero chemical 
potential or net charge density the BCS
mechanism can lead to the condensation of Cooper pairs of fermions, which will
be discussed elsewhere.}    

The evolution equations (\ref{Fscalar}) and (\ref{rhoscalar}) for the scalar
correlators are second-order in time and one needs to specify initial 
conditions
for the propagators and their time derivatives. As for the fermions, 
the initial
conditions for the scalar spectral function is completely specified by the
field commutation relations (cf.~Eq.~(\ref{initialrhoscalar})). 
For the scalar two-point function $F_\phi$ we consider 
(cf.~also Refs.~\cite{Berges:2001fi,Aarts:2001qa,Berges:2002cz})
\bea
 F_\phi(t,t',p)|_{t=t'=0} & = & \frac{1}{\epsilon_0(p)}
 \Big[n_0(p)+\frac{1}{2}\Big] \, , \nonumber\\
 \partial_{t}F_\phi(t,t',p)|_{t=t'=0} & = & 0 \, ,
\label{initialFphi}\\
 \partial_{t}  \partial_{t'} F_\phi(t,t',p)|_{t=t'=0} & = & 
 \epsilon_0(p)\,\Big[n_0(p)+\frac{1}{2}\Big] \, , \nonumber
\eea
with an initial particle number distribution $n_0(p)$ and initial mode energy
$\epsilon_0(p)$ (see \ref{sec:quasiparticle}).

It is instructive to consider for a moment the solution of 
the {\em free field} equations,
which can be obtained from Eqs.\ (\ref{rhoV0eom})--(\ref{rhoVeom}) and
(\ref{FV0eom})--(\ref{FVeom}) by neglecting the memory integrals on their RHS.
The solution of the fermionic free field equations with the above initial
conditions reads
\bea
 F_V(t,t',p)&=&\left(\frac{1}{2}-n_0^{f}(p)\right)\,\cos[p(t-t')] \, ,\nonumber\\
 F_V^0(t,t',p)&=& - i\,\left(\frac{1}{2}- n_0^{f}(p)\right)\,\sin[p(t-t')]\, \, \quad\,\,  \nonumber
\eea
and for the spectral functions one obtains
\beq
 \rho_V(t,t',p)= \sin[p(t-t')] \quad , \qquad
 \rho_V^0(t,t',p)= i \cos[p(t-t')]\, . \nonumber
\eeq
One observes that each mode of the {\em equal-time} correlator 
$F_V(t,t,p)$ is strictly 
conserved in the absence of the memory integrals. 
Since this correlator is directly 
related to particle number (see \ref{sec:quasiparticle}), this means that the 
latter is conserved mode by mode in this approximation. Although this 
is expected in the free field limit, this is of course not the case in the 
fully interacting theory. We emphasize that such additional 
conservation laws do not only appear in the free field limit, but are a 
property of mean-field-type approximations, 
which include local corrections to the bare mass, but neglect the 
scattering contributions described by the memory integrals on the RHS of 
Eqs.~(\ref{rhoV0eom})-(\ref{FVeom}). In these approximations, the existence 
of this infinite number of spurious conserved quantities 
prevents the system to 
approach the thermal equilibrium limit at late times. 
This aspect has been discussed in detail in the context of scalar field 
theories in Refs.~\cite{Berges:2001fi,Bettencourt:1997nf}. It is therefore
crucial to go beyond such ``Gaussian'' approximations in order to correctly 
describe in particular the late-time evolution of the system in the 
interacting theory.

\section{Numerical implementation}
\label{sec:numerics}

We numerically solve the evolution equations
(\ref{rhoV0eom})--(\ref{FVeom}) and (\ref{rhoscalar})--(\ref{Fscalar}), 
together with the self-energies 
Eqs.\ (\ref{selfscalarrho})--(\ref{selffermionF}).
The structure of the fermionic equations is reminiscent of the 
form of classical
canonical equations. In this analogy, $F_V(t,t')$ plays the role 
of the canonical 
coordinate and $F_V^0(t,t')$ is analogous to the canonical 
momentum. This suggests 
to discretize $F_V(t,t')$ and $\rho_V(t,t')$ at $t-t'=2n a_t$ (even) and 
$F_V^0(t,t')$ and $\rho_V^0(t,t')$ at $t-t'=(2n+1)a_t$ (odd) time-like lattice 
sites with spacing $a_t$. This is a generalization of the ``leap-frog'' 
prescription for temporally inhomogeneous two-point functions. This implies
in particular that the discretization in the time direction is coarser
for the fermionic two-point functions than for the bosonic ones. This 
``leap-frog'' prescription may be easily extended to the memory integrals 
on the RHS of Eqs.\ (\ref{rhoV0eom})--(\ref{FVeom}) as well.

We emphasize that the discretization does not suffer from the
problem of so--called fermion doublers \cite{montvay}. The spatial 
doublers do not appear since (\ref{rhoV0eom})--(\ref{FVeom})
are effectively second order in $\vec{x}$-space. Writing the equations
for $\vec F_V(t,t',\vec x)$ and $\vec \rho_V(t,t',\vec x)$
starting from (\ref{rhoV0eom})--(\ref{FVeom}) one realizes that instead of
first order spatial derivatives there is a Laplacian appearing.
Hence we have the same Brillouin zone for the fermions and scalars.
Moreover, time-like doublers are easily avoided by using a sufficiently 
small stepsize in time $a_t/a_s$.

The fact that Eqs. (\ref{rhoV0eom})--(\ref{FVeom}) and
(\ref{rhoscalar})--(\ref{Fscalar}) contain memory integrals makes numerical
implementations expensive. Within a given numerical precision it is typically
not necessary to keep all the past of the two-point functions in the memory.
A single PIII desktop workstation with
2GB memory allows us to use a memory array with 470 timesteps (with 2 temporal
dimensions: $t$ and $t'$). We have checked for the presented runs that a 30\%
change in the memory interval length did not alter the results.  For a typical
run 1-2 CPU-days were necessary.

The shown plots are calculated on a $470\times470\times32^3$ lattice.
(The dimensions refer to the $t$ and $t'$ memory arrays and the momentum-space
discretization, respectively.) By exploiting the spatial symmetries described 
in Section~\ref{sec:symmetries} the memory need could be reduced by a factor
of~30.
We have checked that the infrared cutoff is well below any other mass scales
and that the UV cutoff is greater than the mass scales at least by a factor of
three.

To extract physical quantities we follow the time evolution
of the system for a given lattice cutoff up to late times and 
measure the renormalized scalar mass $m$ which is then used to 
set the scale. In the evolution equations we analytically subtract 
only the respective quadratically divergent terms obtained from a 
standard perturbative analysis. 
We emphasize that for the results presented below we use the late-time 
thermal mass to set the scale, and not the vacuum mass for convenience. 

We made runs for a range of couplings $g^2=0.49$ -- $1$ which show very similar
qualitative behavior. Below, we present plots corresponding to $g=1$, for which
the time needed to closely approach thermal equilibrium is the shortest. 
This allows us to obtain an accurate thermalization with the lowest numerical
cost.

\section{Far-from-equilibrium dynamics}
\label{Sectfarfromequilibrium}

\begin{figure}[t]
\begin{center}
\epsfig{file=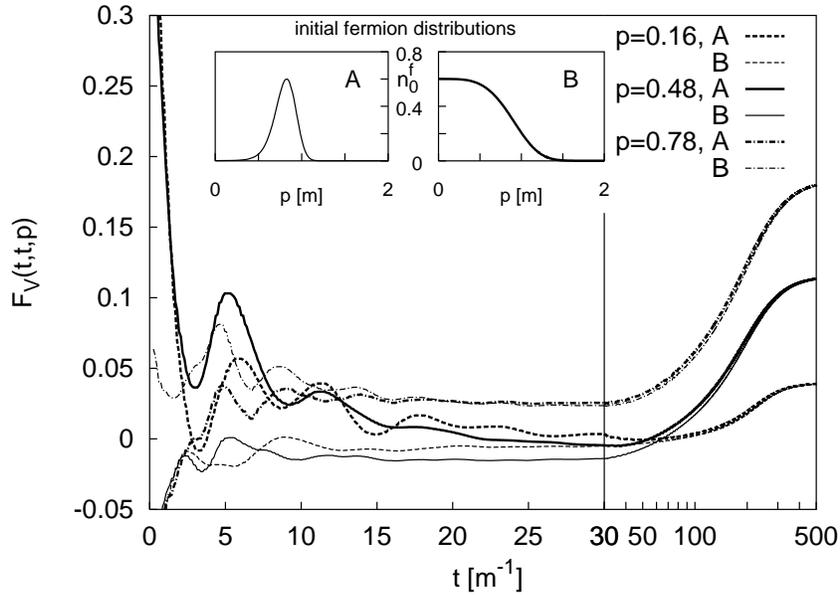,width=11.5cm}
\end{center}
\vspace*{-0.5cm}
\caption{The time evolution of the fermion two-point 
function $F_V(t,t;p)$ for three values of the
momentum $p$, in units of the renormalized scalar thermal mass $m$. 
The evolution is shown for two very different initial conditions 
with the {\em same} initial energy density. One observes that
the dynamics becomes rather quickly insensitive to the initial 
distributions displayed in the insets -- much before the modes 
settle to their final values. The long-time behavior is shown on a 
logarithmic scale for $t \ge 30 m^{-1}$.}
\label{lostinitialfig}
\end{figure}
In Fig.~\ref{lostinitialfig} we present the time evolution 
of the fermion equal-time two-point function $F_V(t,t;p)$ for three 
momenta $p$. Results are given for two very different
initial particle number distributions, which are displayed 
in the insets (see also Eqs. (\ref{initialFV})--(\ref{initialFphi})). 
The (conserved) energy density is taken to be the same for both runs. 
In this case, since thermal equilibrium is uniquely specified by the 
value of the energy density, the correlator modes should approach 
universal values at late times if thermalization occurs. 

\begin{figure}[t]
\begin{center}
\epsfig{file=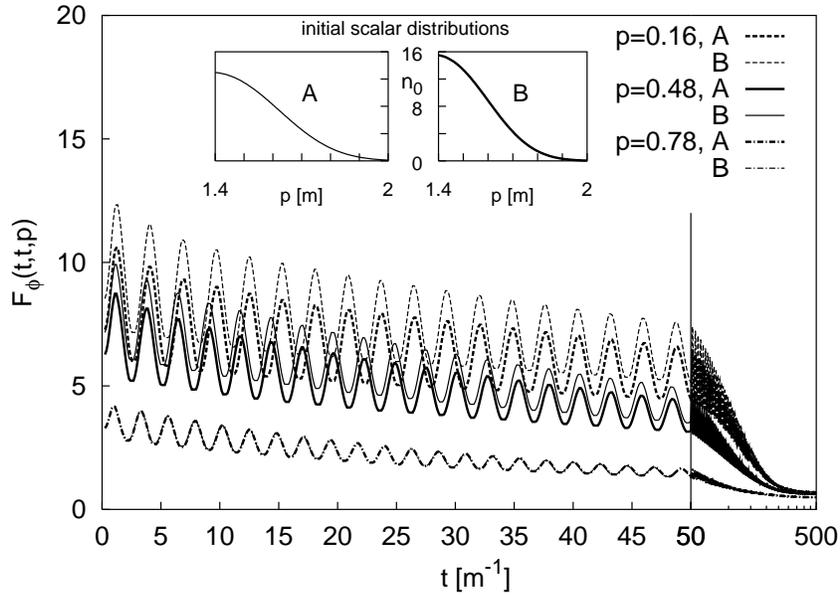,width=11.5cm}
\end{center}
\vspace*{-0.5cm}
\caption{The same as in Fig.~\ref{lostinitialfig} but for the bosonic 
two-point function $F_\phi(t,t;p)$ for three different momenta.
The initial particle number distributions for the two runs denote
by ``A'' and ``B'' are displayed in the insets. For the employed
parameters one observes that, in the scalar sector, the
time needed to become effectively insensitive to the initial 
distributions is comparable to the time scale describing the approach 
to the universal late-time value.}
\label{scalarlostinitialfig}
\end{figure}

It is striking to observe from Fig.~\ref{lostinitialfig} 
that after a comparably short time, much before the correlation 
modes reach their late-time values, the dynamics becomes rather 
insensitive to the details of the initial conditions: for a given momentum,
the curves corresponding to the two different runs come
close to each other rather quickly. During the slow 
subsequent evolution the system is still far away from 
equilibrium before the approach to the late-time values sets in.
From both runs one observes that the characteristic time needed to
effectively loose the information about the details of the initial conditions
is much shorter than the time needed to approach the late-time
result. Moreover, the late-time values are found to be universal in the
sense that the different runs agree with each other to very good 
precision.     

Fig.~\ref{scalarlostinitialfig} shows the corresponding behavior of the
scalar correlator modes $F_\phi(t,t;p)$. The respective initial
particle number distributions in the scalar sector for the two runs
are given in the inset. For the two different runs the modes having the 
same momenta approach each other rather slowly as compared to the fermionic 
sector. However, they reach their final values on a time scale which is 
comparable to that observed for the fermions in Fig~\ref{lostinitialfig}.

\begin{figure}[t]
\begin{center}
\epsfig{file=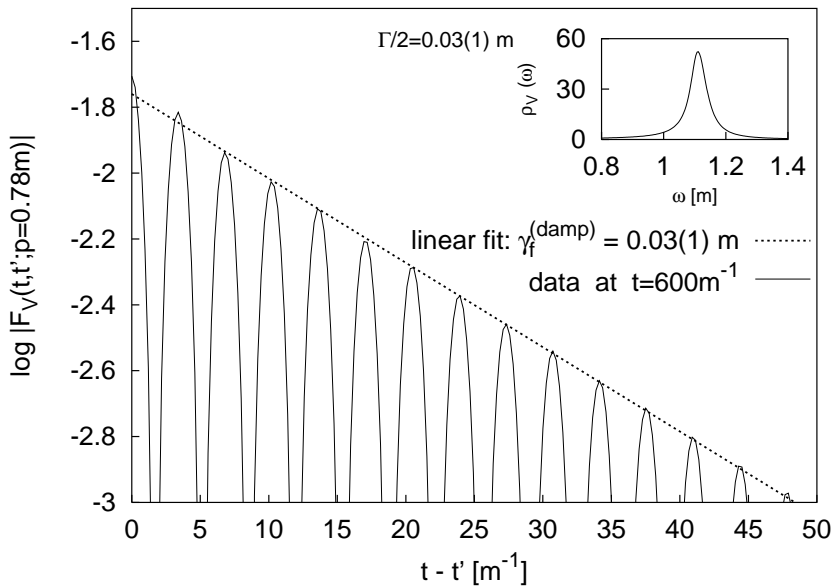,width=11.5cm}
\end{center}
\vspace*{-0.5cm}
\caption{The fermionic two-point function $F_V(t,t';p)$ as a function
of $t-t'$ at late time $t=600\, m^{-1}$. One observes a very good
agreement with an exponential behavior. The rate is well described by the
width of the corresponding spectral function in frequency space
as shown in the inset (cf.\ the text for details).}
\label{spectralfig}
\end{figure}
To characterize these time scales in more detail, we
consider in Fig.~\ref{spectralfig} the unequal-time two-point 
function $F_V(t,t';p)$. As expected, if the system is to become insensitive 
to the details of the initial conditions, we observe that the correlation 
between some time $t$ and another time $t'$ is suppressed for 
sufficiently large $t-t'$. We note that the oscillation envelope of 
$F_V(t,t',p)$ can be well described in terms of an exponential for 
sufficiently late times. The corresponding damping rate approaches 
a constant value. To estimate the asymptotic rate we show
in Fig.~\ref{spectralfig} the unequal-time two-point function
$F_V(t,t';p)$ as a function of $t-t'$ for the late time
$mt = 600$. The fit to an exponential 
yields the damping rate $\gamma^{\rm (damp)}_f(p=0.78m) = 0.03(1)\, m$.  We
find a moderate momentum dependence of this rate with $\gamma^{\rm (damp)}_f(0)
= 0.067(1)m$ and $\gamma^{\rm (damp)}_f(0) \gtrsim \gamma^{\rm (damp)}_f(p >
0)$.

We emphasize that the rate $\gamma^{\rm (damp)}_f(p)$ can be 
related to the width of the Fourier transform of the spectral function
with respect to the time difference $t-t'$. In principle, the latter involves 
an integration over an infinite time interval. However, since we 
consider an initial-value problem for finite times we know 
$\rho_V(t,t',p)$ only on a finite interval. To overcome
this problem, we fit the data for $\rho_V(t,t',p)$ by a $7$--parameter
formula that is capable to account for the observed oscillations and damping, 
but which is more general than the usual 2-parameter Breit-Wigner 
formula. We perform the Fourier transformation on the extrapolated 
data. The resulting function $\rho_V(\omega,p)$ is displayed as a function
of frequency $\omega$ in the inset of Fig.~\ref{spectralfig}. One clearly 
observes a nonzero width of the spectral function, the value of which may 
be obtained from a fit to a Breit-Wigner formula. By doing so, we obtain 
a very good agreement of the damping rate inferred from the width of the 
spectral function on the one hand and from the linear fit on the log-plot 
for $F_V(t,t';p)$ on the other hand.

We can use the fermion damping rate to quantify the time
scale characterizing the effective loss of the details of the initial
conditions: Comparing with Fig.~\ref{lostinitialfig}, we observe 
that the inverse fermion damping rate at $p\simeq 0$
$\left(1/\gamma^{\rm (damp)}_f = 15(1) m^{-1}\right)$ 
characterizes well the time for which the dynamics becomes rather 
insensitive to the initial distributions. In contrast, we find 
that this time scale does not characterize the late-time
behavior. For the latter, we observe to very good approximation an 
exponential relaxation of each mode $F_V(t,t;p)$ to its universal 
late-time value. Carrying out the measurement for different modes, we
observe that the corresponding rate is almost independent of momentum 
and is given by $1/\gamma^{\rm (therm)}_f=95(5) m^{-1}$. The corresponding
time scale $1/\gamma^{\rm (therm)}_f$ is therefore much larger
than the characteristic damping time $1/\gamma^{\rm (damp)}_f$.

A similar analysis can be performed for the scalar sector.
Here we find at $p\simeq 0$ the corresponding values
$1/\gamma^{\rm (damp)}_\phi = 50(5)\,m^{-1}$ 
and $1/\gamma^{\rm (therm)}_\phi = 90(5) m^{-1}$.
One observes that although the damping rates for fermions 
and bosons are very different, the respective 
thermalization rates are rather similar. 
For the scalars the thermalization rate is larger than the damping rate,
as is observed above for the fermions.
However, the difference is much less pronounced for the 
scalars. Similar studies in $1+1$ dimensional
quantum \cite{Berges:2001fi} and classical \cite{Aarts:2001wi} scalar theories 
typically find a substantial difference between damping and thermalization 
rates. However, this may be a consequence of the stringent phase-space 
restrictions and therefore particular to $1+1$ dimensional systems.

We finally note that approximate rates describing the early-time
behavior may be determined by an exponential fit to the functions
$F_\phi(t,0)$, $F_V^0(t,0)$ and $F_V(t,0)$ in a finite time interval. 
These rates depend on time and approach the late-time values 
given above.  At early times we observe an approximate exponential 
damping with a rate about twice as big as the late-time value for the 
fermions, and about half the late-time rate for the scalars.

\section{Approach to quantum thermal equilibrium}
\label{sec:statistics}

In the previous section, we have seen that the out-of-equilibrium 
evolution of the system leads to a universal late-time behavior,
uniquely characterized by the initial energy density. We now
analyze in detail if quantum thermal equilibrium, characterized
by Bose-Einstein and Fermi-Dirac statistics, is 
approached\footnote{We emphasize 
that thermal equilibrium cannot be reached on a fundamental
level from time-reversal invariant evolution equations at any
finite time. The results demonstrate that thermal equilibrium 
can be approached very closely at sufficiently late time,
without again deviating from it for practically accessible times.
For a more detailed discussion of this aspect see
Ref.~\cite{Berges:2001fi}.}.
In thermal equilibrium, the spectral function
and the statistical two-point function are not independent of each
other, but are related by the fluctuation-dissipation relation. The latter
is an exact relation, which can be stated in $4$--dimensional Fourier space 
as~\cite{Berges:2001fi,Aarts:2001qa}
\beq
F^{(\rm eq)}_{\phi}(\omega,\vec{p}) = - i\left(n^{BE}(\omega) + 
\frac{1}{2}\right)
\rho^{(\rm eq)}_{\phi}(\omega,\vec{p}) \, ,
\label{eq:fluctdiss}
\eeq 
for the scalar correlators, where $n^{BE}(\omega) = 1/[\exp (\omega/T) -1]$
denotes the Bose--Einstein distribution function. The frequency $\omega$ 
is the Fourier conjugate of $t-t'$ (in thermal equilibrium, 
time-translation invariance implies that two-point functions
only depend on $t-t'$). The value of the
temperature $T$ is determined by the energy density of the system. The 
fluctuation-dissipation relation for the fermion correlators 
is given by the equivalent expression with the replacement 
$(n^{BE}(\omega) + \frac{1}{2}) \to (n^{FD}(\omega) - \frac{1}{2})$
in Eq.~(\ref{eq:fluctdiss}) and the Fermi--Dirac distribution
$n^{FD} = 1/[\exp (\omega/T) + 1]$. The same relations
hold as well for the statistical and spectral components of the
self-energies in thermal equilibrium. They provide an unambiguous 
way to extract the distribution functions, which are specific for 
quantum thermal equilibrium.
Out of equilibrium, the spectral and statistical two-point functions are 
completely independent in general. However, if thermal equilibrium
is approached at late times, they become related by the 
fluctuation-dissipation relations.

\begin{figure}[t]
\begin{center}
\epsfig{file=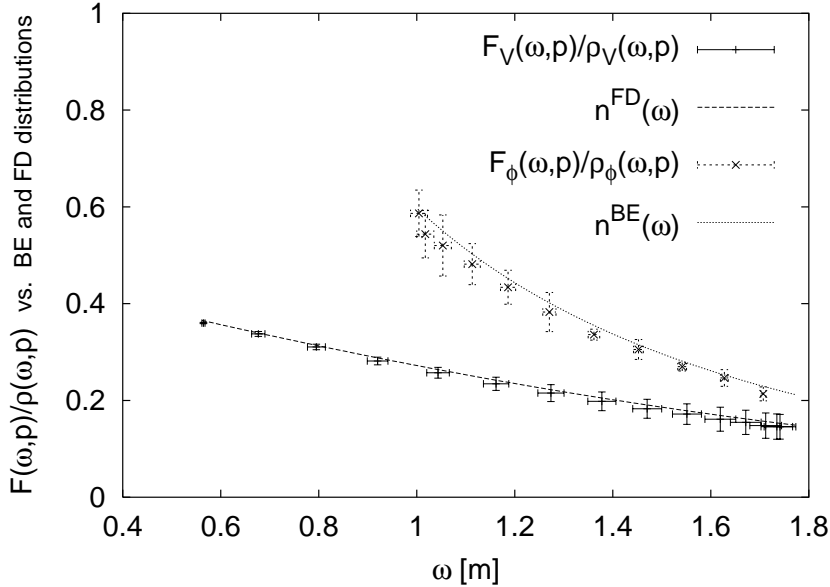,width=11.5cm}
\end{center}
\vspace*{-0.6cm}
\caption{The late-time ratio of the statistical two-point function  
and the spectral function in frequency space, both for
fermions ($F_V/\rho_V$) and for scalars ($F_\phi/\rho_\phi$).  
In thermal equilibrium the quotient corresponds to the Bose-Einstein
(BE) distribution function for scalars and to the Fermi-Dirac (FD) 
distribution for fermions -- independently of any assumption
on a quasi-particle picture (see Eq.\ (\ref{eq:fluctdiss})). 
The BE/FD distributions are displayed by the continuous curves 
parametrized by the same temperature. The value of the latter, $T=0.94m$,
is actually not fitted but has been taken from the inverse slope 
of Fig.~\ref{compslopefig}. This shows the correspondence with the 
quasi-particle picture described in the text.}
\label{fig:fluctdiss}
\end{figure}

To extract the answer about the late-time distributions, 
without relying on any assumptions, we consider the statistical
and spectral correlators in
Wigner coordinates. For this we express $F_{\phi}(t,t';p)$ and 
$\rho_{\phi}(t,t';p)$, as well as the corresponding fermion 
two-point functions, in terms
of the center coordinate $X^0=(t+t')/2$ and the relative
coordinate $s^0=t-t'$ and write
\beq
\label{irho}
\rho_{\phi}(X^0;\omega,p) =  \int_{-2 X^0}^{2 X^0}
{\rm d}s^0\, e^{i \omega s^0}\, 
\rho_{\phi}(X^0+s^0/2,X^0-s^0/2;p) \, , 
\eeq  
and equivalently for the other correlators.
Since we consider an initial-value problem,
the time integral over $s^0$ is bounded by $\pm 2 X^0$ (cf.~also the 
detailed discussion in Ref.~\cite{Aarts:2001qa}).
If thermal equilibrium is approached for sufficiently large $X^0$,
then the correlators do no longer depend on $X^0$ and a Fourier
transform with respect to $t-t'$ can be performed to very good 
approximation (cf.~the discussion in Sect.~\ref{Sectfarfromequilibrium}). 
The distribution functions may then be extracted from
the quotient of the Wigner transformed two-point functions. 
In Fig.~\ref{fig:fluctdiss} we show the respective ratios 
at late times. Both functions are in good agreement 
with the equilibrium distributions $n^{FD}(\omega)$ and $n^{BE}(\omega)$, 
respectively (cf.~Eq.~(\ref{eq:fluctdiss})).
We emphasize that the displayed continuous curves are no separate fits. 
They are the Bose-Einstein and Fermi-Dirac distribution functions
parametrized by the same temperature. In particular, the value for the
temperature is not fitted but has been extracted from the inverse 
slope parameter as explained below.

\begin{figure}[t]
\begin{center}
\epsfig{file=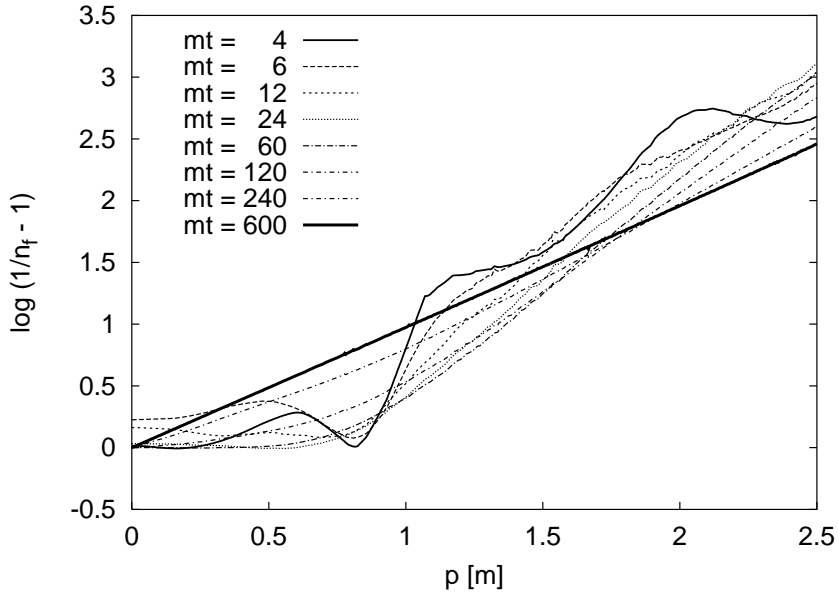,width=11.5cm}
\end{center}
\vspace*{-0.6cm}
\caption{The time-dependent fermion quasi-particle distribution 
$n_f(t,p)$ as a function of mode energy $p$ for various times $t$. 
We have plotted the inverse slope function $\log(1/n_f -1)$, which 
reduces to a straight line intersecting the origin when $n_f(t,p)$ 
approaches a Fermi-Dirac distribution. This plot shows the data for 
run ``A'' of Fig.~\ref{lostinitialfig}.}
\label{finvslopefig}
\end{figure}

We stress that the above procedure to extract the distribution 
functions is independent of any assumption about a quasi-particle 
picture. However, for many practical purposes it is very convenient 
to have an effective description of particle number and mode
energy directly in real time -- without the need of a Fourier transform. 
An efficient description is elaborated in \mbox{\ref{sec:quasiparticle}}. The 
value for the temperature in the thermal equilibrium distributions 
of Fig.~\ref{fig:fluctdiss} has been actually measured  
based on this quasi-particle picture. In \ref{sec:quasiparticle} we define 
the effective particle number and energy to be used, both for the fermionic 
and bosonic fields. For scalar field theories, this quasi-particle 
picture has been successfully employed previously to investigate 
thermalization~\cite{Berges:2001fi,Aarts:2001yn,Aarts:2001qa}.

\begin{figure}[t]
\begin{center}
\epsfig{file=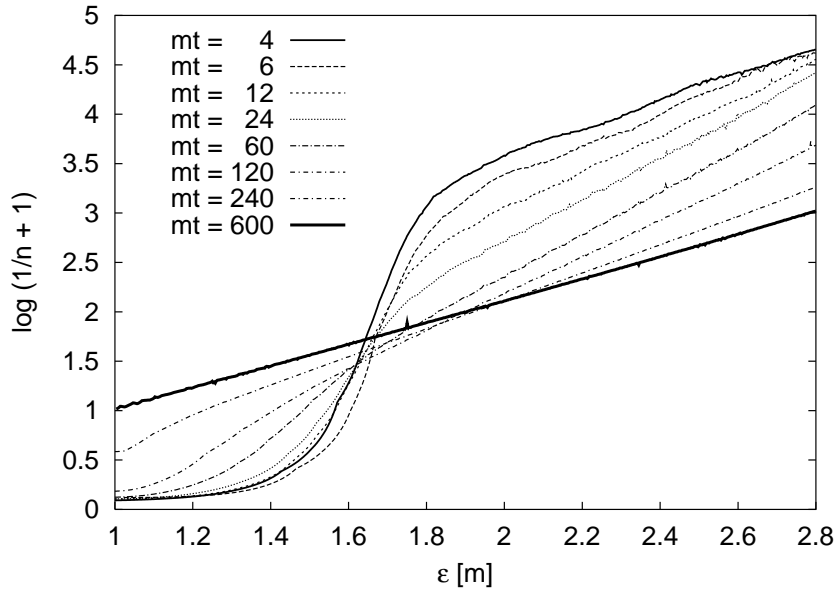,width=11.5cm}
\end{center}
\vspace*{-0.6cm}
\caption{The time-dependent boson quasi-particle distribution 
$n(t,p)$ as a function of mode energy $\epsilon(t,p)$ (see text) 
for various times. In this case the inverse slope function is 
$\log(1/n + 1)$, which reduces to a straight line in case of a 
Bose-Einstein distribution. This plot shows the data for run ``A''
of Fig.~\ref{scalarlostinitialfig}.}
\label{sinvslopefig}
\end{figure}

In Figs.~\ref{finvslopefig}~and~\ref{sinvslopefig} we show the 
effective quasi-particle number distributions defined as 
\beq
 \frac{1}{2} - n_f(t,p) = F_V(t,t;p)
\label{fermionnumber}
\eeq
for fermions and  
\bea
 \frac{1}{2} + n(t,p) &=& \epsilon(t,p) \, F_\phi(t,t;p) \, , \\
\label{bosonnumber}
 \epsilon(t,p) &=& 
 \left(\frac{\partial_t\partial_{t'}F_\phi(t,t';p)}{F_\phi(t,t';p)}
 \right)_{t=t'}^{1/2}
\label{bosonenergy}
\eea
for scalars, where $\ep(t,p)$ is the quasi-particle mode energy  
as discussed in \ref{sec:quasiparticle}.\footnote{Note that because 
of chiral symmetry there is no mass term present for the fermions 
and the corresponding quasi-particle
mode energy is simply $p$.}  The curves correspond to the initial conditions of
run ``A'' shown in Figs.~\ref{lostinitialfig}~and~\ref{scalarlostinitialfig}.
\begin{figure}[t]
\begin{center}
\epsfig{file=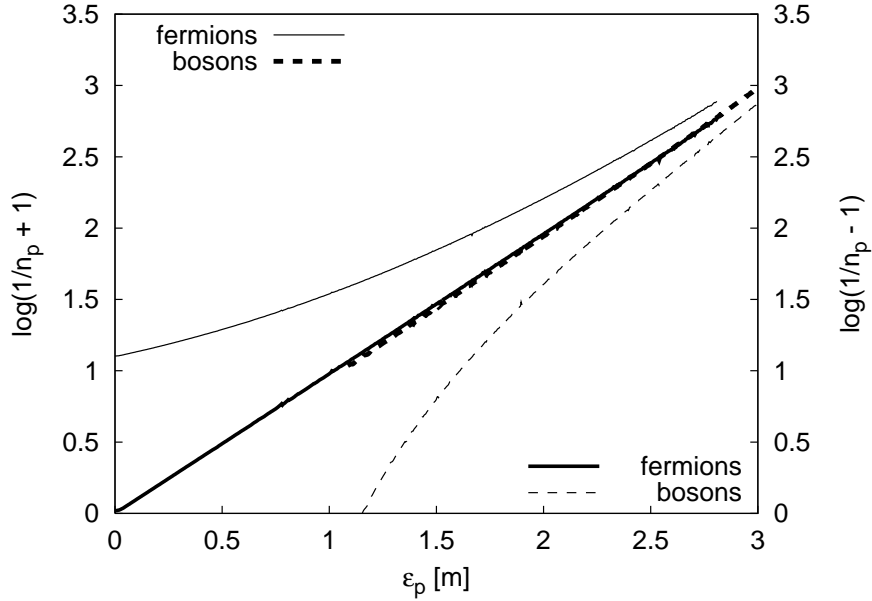,width=11.5cm}
\end{center}
\vspace*{-0.6cm}
\caption{Boson and fermion quasi-particle distributions at late times
as a function of mode energy ($n_p\equiv n(t,p)$ and 
$\epsilon_p\equiv \epsilon(t,p)$ for bosons
and $n_p\equiv n_f(t,p)$ and $\epsilon_p\equiv p$ for fermions). 
Both inverse slope functions employed in 
Figs.~\ref{finvslopefig}~and~\ref{sinvslopefig} have been applied here 
in order to demonstrate that the fermions clearly do not follow a Bose-Einstein
distribution, and the scalars cannot be characterized by a 
Fermi-Dirac distribution. In contrast, for
the respective correct statistics the curves lie on top of each other,
showing that both fermions and scalars are  
described by the same inverse slope parameter. The value of the latter is used
in Fig. \ref{fig:fluctdiss} as the temperature for the distributions
$n^{FD}(\omega)$ and $n^{BE}(\omega)$.}
\label{compslopefig}
\end{figure}
One observes how the effective fermion and boson particle numbers 
change with time, the former approaching a Fermi--Dirac and the latter 
a Bose--Einstein distribution.
To emphasize this point, we plot the corresponding ``inverse slope functions'' 
$\log(1/n_f -1)$ and $\log(1/n + 1)$, which reduce to straight lines for 
Fermi--Dirac and Bose--Einstein distributions respectively. The associated 
inverse slopes correspond to the temperature of the thermal equilibrium
distributions. We see in 
Figs.~\ref{finvslopefig}~and~\ref{sinvslopefig} that both inverse slope 
functions approach straight lines at late times. The associated temperatures 
for fermions and bosons are independent of time and agree very well with 
each other, as shown in Fig.~\ref{compslopefig}. For comparison, 
we display in Fig.~\ref{compslopefig} the fermion inverse slope function
evaluated with the bosonic effective particle number, and vice versa. This
illustrates the degree of sensitivity of the inverse slope functions to the
different statistics and in turn the degree of precision with which we are
able to probe thermalization.

\section{\label{sec:conclusions} Conclusions}

In this paper we have discussed the far-from-equilibrium 
dynamics and subsequent thermalization of a system of coupled 
fermionic and bosonic quantum fields. We solved the nonequilibrium
dynamics beyond mean-field type approximations by calculating the 
complete lowest non-trivial order in a systematic coupling expansion 
of the 2PI effective action, which includes direct scattering 
as well as memory and off-shell effects. To our knowledge this is 
the first time that such a 
calculation is performed without further approximations. As a result, 
we show that, for various far-from-equilibrium initial conditions,
the late-time behavior is universal and uniquely determined by the value 
of the initial energy density. Moreover, we are able to probe the
approach to quantum thermal equilibrium, characterized by 
the emergence of Fermi--Dirac and Bose--Einstein distribution
functions, with high accuracy.
We emphasize that in the present calculation, besides the limitations
of a coupling expansion, there is no other 
input than the dynamics dictated by the considered quantum field theory for 
given nonequilibrium initial conditions.

This work can be extended in many directions. Most of the equations we 
derive are also valid in the phase with spontaneous symmetry breaking.
Combined with earlier work on scalar theories 
\cite{Berges:2001fi,Aarts:2002dj,Berges:2002cz}, 
this provides a description of the 
nonequilibrium dynamics of the linear $\sigma$-model for QCD with 
two quark flavors. The model has served for many years as a valuable 
testing ground for ideas on the equilibrium phase structure of low 
energy QCD at nonzero temperature and density. With the present 
techniques a quantitative understanding of the out-of-equilibrium 
physics of this model is within reach.

\subsection*{Acknowledgment}
We thank C.~Wetterich for many discussions and collaboration
on related work. We are also grateful to W.~Wetzel for his continuous
support with computers. Sz.~B.\ acknowledges the hospitality of the Institute 
f\"ur Theoretische Physik, Heidelberg. His work was supported by 
the short-term scholarship program of the DAAD.

\appendix
\section{\label{sec:quasiparticle}Effective particle number and energy}

A particle number can only be strictly defined in the presence of a conserved
charge. However, for physical interpretation it is often convenient to define
an effective particle number even when there is no conserved charge. 
In particular, besides a total particle number it is often useful
to have a definition of an effective particle number per (momentum) mode. 
The latter is typically not conserved in an interacting theory, and
in the context of thermalization one would like to find a time-dependent
particle number distribution,
which allows one to observe a Bose-Einstein or Fermi-Dirac distribution at
sufficiently late times\footnote{We emphasize, however, that the definition
of an effective particle number distribution is not necessary to
analyse the approach to quantum thermal equilibrium, as is discussed
in Sect.~\ref{sec:statistics}.}. Of course, the notion of an effective 
particle number is typically only meaningful if the relevant degrees 
of freedom or ``quasi-particles'' are weakly interacting. 
    
There are many different ways of how one can introduce the notion of 
an effective particle number (for a recent discussion 
see e.g.~Ref.~\cite{Garbrecht:2002pd} and references therein). 
For example, one can define it as the 
average energy per mode divided by the energy of the corresponding mode. 
In an interacting theory, this procedure can be ambiguous because 
the expression of the total energy receives contributions from interactions 
and the average energy per mode is not uniquely defined. 
In this appendix, we discuss an elegant way to circumvent this difficulty. 
More importantly for our purposes, the procedure leads to a definition of an 
effective particle number density, which indeed allows one to 
directly observe the emergence of a Fermi-Dirac or 
Bose-Einstein distribution from the nonequilibrium dynamics
for the theory considered in this paper. 
In particular, it can be explicitely 
shown that the effective particle numbers are always positive 
and, for the fermions, smaller than one (see \ref{sec:bounded}).
Applying the present construction to the case of neutral scalar
fields, we recover the particle number definition used in previous studies 
to exhibit thermalization in purely scalar field 
theories~\cite{Aarts:2001qa,Berges:2001fi}.

\vspace{.3cm}
\noindent
{\bf Fermions}
\vspace{.3cm}

\noindent
We start by considering the 
case of charged fields and construct the effective particle number  
from the conserved current generated by the $U(1)$ symmetry.\footnote{As 
we shall see below for scalar fields, the expression for the effective 
particle number one obtains in this way can be directly applied to the case of
neutral fields as well.}
The associated $4$--current for each given flavor is 
$\sim \bar\psi \gamma^\mu \psi$. Fourier transformed with respect to spatial
momenta, the expectation value of 
the latter can be written as 
$J_f^\mu (t,p) = \tr [\gamma^\mu D^<(t,t,p)]$, 
where the subscript stands for fermions (cf.~Eq.~(\ref{Dbs})). 
In terms of the equal-time statistical two 
point function, its temporal and spatial components read:\footnote{The constant
factor in the temporal component comes from the fact that we define the current
without the standard normal ordering.}
\bea
 J_f^0 (t,p) & = & 2 \, [1 - 2\,F_V^0(t,t;p) ] \, ,\nonumber \\
 \vec J_f (t,p) & = & -4\bv \, F_V(t,t;p)\, .\nonumber
\eea
A nice property is that the above 
expressions in terms of the full correlators 
do not contain any explicit dependence 
on the interaction part. 
In order to obtain an effective particle number,
we want to identify these expressions with the corresponding ones in a 
quasi-particle description with free-field expressions. These
are given by:\footnote{Here, we have explicitly used our 
assumptions of parity symmetry and rotational invariance
(cf.~Sect.~\ref{sec:symmetries}), which imply that the 
different spin states contribute the same, therefore the factor of two.}
\bea
 J_{f}^{0\, ({\rm QP})} (t,p) & = & 2 \, [1+Q_f(t,p)] \, ,\nonumber \\
 \vec J_{f}^{\,({\rm QP})} (t,p) & = & -2 \bv \,[1- 2N_f(t,p)] \, ,\nonumber
\eea
where $Q_f(t,p)=n_f-\bar{n}_f$ is the difference between particle and 
anti-particle effective number densities and 
$N_f(t,p)=(n_f+\bar{n}_f)/2$ is their half-sum. The 
physical content of these expressions is simple: the temporal component 
$J^0$ directly represents the net-charge density per mode 
$Q_f(t,p)$, whereas the spatial 
part $\vec J$ is the net current density per mode and is therefore sensitive to
the sum of particle and anti-particle number densities.
Identifying the above expressions, we define
\bea
 \frac{1}{2}\,Q_f(t,p) &=& -F_V^0(t,t;p) \, ,\label{genericcharge} \\
 \frac{1}{2}-N_f(t,p) &=& F_V(t,t;p) \, .\label{genericparticle}
\eea
Of course, these expressions are only meaningful for the case that 
the interacting theory is well-described by a quasi-particle picture. 
The important point is that the above procedure allows one to construct 
an effective particle number density without knowing a priori which 
part of the interaction is to be considered as the ``dressing'' of  
the quasi-particles and which part describe their ``residual'' interactions.
We note that the equal-time two-point functions on the RHS of these equations
are real by definition (see Eq.~(\ref{CP})). Moreover, 
using the anticommutation 
relation for the fermion fields, it is shown in
\ref{sec:bounded} that these definitions 
always satisfy $0 \le N_f(t,p) \le 1$ and $-1 \le Q_f(t,p) \le 1$.
These properties are important for the above definitions to be physically 
meaningful. 

Introducing a non-vanishing net charge density per mode was useful for the
above general construction. However, in the present paper, we consider only
$CP$--invariant systems, which imply that the latter should vanish. Indeed,
we see from Eqs.~(\ref{CP}) that the requirement of $CP$--invariance imply
that our above definition of net charge density per mode vanishes
identically for all times and for all modes. Therefore, the effective 
particle and anti-particle numbers are equal and we have
\bea
 Q_f(t,p) &=&0 \, ,\nonumber\\
 \frac{1}{2}-n_f(t,p) &=& F_V(t,t;p) \, .\nonumber
\eea
where $n_f(t,p)$ is the effective particle number.

\vspace{.3cm}
\noindent
{\bf Bosons}
\vspace{.3cm}

\noindent
Following the same lines as above, let us consider for a moment the case
of a single charged scalar field $\phi$.
The $4$-current associated the corresponding $U(1)$ symmetry is 
$\sim i[\phi^\dagger (\partial^\mu \phi) - (\partial^\mu \phi^\dagger) \phi]$ 
and, as for the case of fermions, 
its expectation value has a simple expression 
in terms of the equal-time statistical two-point 
function of the charged field. 
In momentum space, it reads:
\bea
 J_b^0 (t,p) & = &  i (\partial_t-\partial_{t'}) \,F_\phi(t,t';p) |_{t'=t} -1 
\, ,
 \nonumber \\
 \vec J_b (t,p) & = & 2 \,\bp\, F_\phi(t,t;p)\, ,\nonumber
\eea
where, as before, the constant contribution comes from the fact that we define
the current without the usual normal ordering. Here, the statistical two-point
function for the charged scalar field is defined as the expectation value of
the anticommutator of two fields operators: 
$F_\phi(t,t';p)=\mbox{$\frac{1}{2}
\langle[\phi(t,\bp),\phi^\dagger(t',\bp)]_+\rangle$}$. It has the symmetry
property $F_\phi(t,t';p)=F_\phi^*(t',t;p)$, so that the above expressions are real.
The corresponding quasi-particle expressions read:
\bea
 J_{b}^{0\,({\rm QP})} (t,p) & = & Q_b(t,p)-1 \, ,\nonumber \\
 \vec J_{b}^{\,({\rm QP})} (t,p) & = & \frac{\bp}{\epsilon(t,p)} \, 
[1+2N_b(t,p)] 
 \, ,\nonumber
\eea
where $Q_b(t,p)$ and $N_b(t,p)$ have the same meaning as before in terms
of effective particle and anti-particle number densities and where
$\epsilon(t,p)$ is the quasi-particle energy. We therefore define:
\bea 
 Q_b(t,p) &=& i (\partial_t-\partial_{t'}) \,F_\phi(t,t';p) |_{t'=t} \, ,
\label{app_scalarcharge} \\
 1+ 2N_b(t,p) &=& 2 \, \epsilon(t,p) \, F_\phi(t,t;p)\, .
\label{app_scalarnumber}
\eea
Note that the RHS of both expressions are real quantities, as they should if
the LHS are to be interpreted as charge and quasi-particle number densities
respectively.

It remains to define the effective quasi-particle 
energy $\epsilon(t,p)$.\footnote{Note 
that this was not necessary for the fermionic case, because of our assumption
of chiral symmetry, which prevents an effective mass term. 
In fact, one can repeat the following 
argument to obtain for the fermion effective quasi-particle 
energy $\epsilon_f (t,p) = p$, as employed above.} For this purpose, 
we use the free-field like expression for the average energy per mode, 
which in the case of a single charged scalar field can be written as
$$
 \partial_t \partial_{t'} F_\phi(t,t';p) |_{t'=t} +
 \epsilon^2(t,p) F_\phi(t,t;p) \equiv \epsilon(t,p) \, \Big[2n(t,p)+1\Big] \, 
$$
in terms of the statistical two-point function.
Therefore, one obtains for the effective quasi-particle energy:
\beq
\label{app_scalarenergy}
 \epsilon^2(t,p) = \left(\frac{\partial_t\partial_{t'}F_\phi(t,t';p)}{F_\phi(t,t';p)}
 \right)_{t=t'} \, .
\eeq
It is straightforward to show that the combination of equal-time correlators
appearing on the RHS of (\ref{app_scalarenergy}) is indeed positive. Moreover, 
using commutation relations for the scalar field, one can show that the
effective particle number $N_b(t,p)$, as given by
Eqs.~(\ref{app_scalarnumber}) and (\ref{app_scalarenergy}), is a positive
quantity (see~\ref{sec:bounded}).  
We note that for the case of
$CP$--invariant systems, the effective charge density per mode
(\ref{app_scalarcharge}) vanishes, as it should.\footnote{This immediately
follows from the behavior of the statistical propagator under
$CP$--transformation: $F_\phi (t,t';p) \longrightarrow F_\phi (t',t;p)$.} 

When dealing with neutral scalar fields, as it is the case in the present
paper, we can use the same formula derived above. Notice that this is
consistent since in that case one has $F_\phi(t,t';p)=F_\phi(t',t;p)$, which
directly implies that $Q_b(t,p) = 0$. Therefore, in the present paper, we use
the definition
\beq
\label{app_neutral}
 \frac{1}{2}+ n(t,p) = \epsilon(t,p) \, F_\phi(t,t;p) \\
\eeq
together with (\ref{app_scalarenergy}) for the boson effective particle number
$n(t,p)$ and mode energy $\epsilon(t,p)$ for each individual scalar 
species~\cite{Aarts:2001qa,Berges:2001fi}.

\section{\label{sec:bounded}}

Here we show that the combinations of equal-time two-point functions  
used in this paper to define effective particle number densities are 
always positive and, for the fermionic case, smaller than one. It is
demonstrated to be a simple consequence of the (anti-)commutation 
relations of field operators.

\vspace{.3cm}
\noindent
{\bf Fermions}
\vspace{.3cm}

\noindent
As a first exercise, we show that for any operators $\varphi$ and 
$\varphi^\dagger$ satisfying the anticommutation relations 
$$
 [\varphi,\varphi^\dagger]_{_{_+}}=1 
$$
and 
$$
 [\varphi,\varphi]_{_+}=[\varphi^\dagger,\varphi^\dagger]_{_+}=0 \, ,
$$
one has
\beq
\label{App_ineqfermion}
 0 \,\le\, \<\varphi^\dagger\,\varphi\> \,\le\, 1 \, ,
\eeq
where the brackets denote an average with respect to any density matrix.
The left inequality above is trivially obtained for example by inserting a
complete sum of states between the operators $\varphi^\dagger$ and $\varphi$.
One obtains a sum of positive quantities which is of course positive. To show
the second inequality, we introduce the hermitian operator $N=\varphi^\dagger
\varphi$. Using the anticommutation relations above, it is easy to see that
$N^2=N$, from which it follows that 
$$
 \Delta n^2 \equiv \Big\<\,\Big(N-\<N\>\Big)^2\,\Big\> = \<N^2\> - \<N\>^2 = 
 \<N\> \Big(1-\<N\>\Big) \, .
$$
It is clear that $\Delta n^2 \ge 0$ and we have just shown that $\<N\> \ge 0$.
One therefore conclude from the above equation that $\<N\> \le 1$, 
as announced.

We now come to our effective particle number densities.  Using
Eqs.~(\ref{genericcharge})
and (\ref{genericparticle}) and recalling that $Q_f\equiv n_f-\bar n_f$ and
$N_f\equiv(n_f+\bar n_f)/2$, we get for the effective particle and
anti-particle number densities:
\bea
 \frac{1}{2}-n_f(t,p)&=&F_V(t,t,p)+F_V^0(t,t,p) 
 = \frac{1}{4} \tr \Big[(\gamma_0 + \bv\cdot\vec\gamma)F(t,t,p) \Big] \, ,
 \nonumber\\
 \frac{1}{2}-\bar{n}_f(t,p)&=&F_V(t,t,p)-F_V^0(t,t,p)
 =-\frac{1}{4} \tr \Big[(\gamma_0 - \bv\cdot\vec\gamma)F(t,t,p) \Big]\, .  
 \nonumber
\eea
In terms of the fermionic field operators $\psi(t,\bp)$ and $\bar\psi(t,\bp)$,
which satisfy the anticommutation relations
$$
 [\psi(t,\bp),\bar\psi(t,\bp)]_{_+}=\gamma_0
$$
and 
$$
 [\psi(t,\bp),\psi(t,\bp)]_{_+}=[\bar\psi(t,\bp),\bar\psi(t,\bp)]_{_+}=0 \, , 
$$
one has 
$$
 F(t,t,p)=\frac{1}{2} \Big\< [\psi(t,\bp),\bar\psi(t,\bp)]_{_-} \Big\> \, .
$$
Therefore, one can rewrite (from now on, we drop the explicit $t$ and $\bp$
dependence):
\bea
 n_f&=&\frac{1}{4} \,\<\bar\psi (\gamma_0 + \bv\cdot\vec\gamma) \psi \> \, , 
 \nonumber\\
 1-\bar n_f&=&\frac{1}{4} \,\<\bar\psi (\gamma_0 - \bv\cdot\vec\gamma) \psi \> 
 \nonumber \, .
\eea
Now we introduce the operator
$$
 \varphi=\frac{\gamma_0 + \bv\cdot\vec\gamma}{\sqrt2}\,\psi \, ,
$$
in term of which,
\beq
\label{App_fpart}
 n_f=\frac{1}{4} \,\sum_{\alpha=1}^4 
 \<\varphi^\dagger_\alpha \varphi_\alpha\> \, ,
\eeq
where we explicitely wrote the sum over Dirac indices.  For a given Dirac
index, it is straightforward to check that
\beq
\label{App_anticom}
 [\varphi_\alpha,\varphi^\dagger_\alpha]_{_{_+}}=
 \Big(1-\gamma_0\, \bv\cdot\vec\gamma\Big)_{\alpha \alpha} = 1
\eeq
and 
$$
 [\varphi_\alpha,\varphi_\alpha]_{_+}=
 [\varphi^\dagger_\alpha,\varphi^\dagger_\alpha]_{_+}=0 \, ,
$$
where we specialized to the Dirac basis to write the last equality of
Eq.~(\ref{App_anticom}).\footnote{This is more convenient, but not necessary.
It is simple to adapt the argument to any basis.} Using (\ref{App_ineqfermion})
for each individual $\alpha$, we conclude from (\ref{App_fpart}) that
$$
 0 \le n_f (t,p) \le 1
$$
for any time $t$ and any momentum $p$. It is straightforward to repeat the above
arguments to show that 
$$ 
0 \le \bar n_f (t,p)  \le 1\, .
$$

\vspace{.2cm}
\noindent
{\bf Bosons}
\vspace{.3cm}

\noindent
For scalars, we first use Eqs.~(\ref{app_scalarenergy}) and
(\ref{app_scalarenergy}) to rewrite
$$
 \frac{1}{2} + n(t,p) = \sqrt{F_\phi(t,t;p) \, 
 [\partial_t \partial_{t'} F_\phi(t,t';p)]_{t'=t}} \, .
$$
Recalling the definition of the statistical propagator in terms of field
operators:
\bea
 F_\phi(t,t;p) &=& \<\phi^\dagger (t,\bp) \, \phi(t,\bp) \> \, , \nonumber\\
 \partial_t \partial_{t'} F_\phi(t,t';p)]_{t'=t} &=&
 \<\Pi^\dagger (t,\bp) \, \Pi(t,\bp) \> \, , \nonumber 
\eea
where $\Pi(t,\bp)$ and $\Pi^\dagger(t,\bp)$ are the canonical momenta conjugate
to $\phi^\dagger(t,\bp)$ and $\phi(t,\bp)$ respectively (e.g.~$\Pi(t,\bp)=\dot\phi^\dagger(t,\bp)$), 
we see that 
$$
 n(t,p) \ge 0 \,\,<=>\,\,
 \<\phi^\dagger \, \phi \> \,
 \<\Pi^\dagger  \, \Pi \> \ge \frac{1}{2} \, ,
$$
where we dropped in the notation 
the explicit time and momentum dependence of field operators. 
The second inequality is nothing but Heisenberg's uncertainty relation, 
a direct
consequence of the equal-time canonical commutation relations of field 
operators \cite{BohmQM}.

\end{document}